\documentstyle[amssymb,epsfig]{mn}

\newbox\grsign \setbox\grsign=\hbox{$>$} \newdimen\grdimen \grdimen=\ht\grsign
\newbox\simlessbox \newbox\simgreatbox \newbox\simpropbox \newbox\wtildebox 
\setbox\simgreatbox=\hbox{\raise.5ex\hbox{$>$}\llap
     {\lower.5ex\hbox{$\sim$}}}\ht1=\grdimen\dp1=0pt
\setbox\simlessbox=\hbox{\raise.5ex\hbox{$<$}\llap
     {\lower.5ex\hbox{$\sim$}}}\ht2=\grdimen\dp2=0pt
\def\simgreat{\mathrel{\copy\simgreatbox}}

\newcommand{\Msun}{\mbox{$M_{\odot}$}}
\newcommand{\Msunpc}{\mbox{$M_{\odot}$~pc$^{-3}$}}
\newcommand{\Mcl}{\mbox{$M_{\rm cl}$}}
\newcommand{\be}{\mbox{\begin{equation}}}
\newcommand{\ee}{\mbox{\end{equation}}}
\newcommand{\Vlim}{\mbox{$V_{\rm lim}$}}
\newcommand{\Rlim}{\mbox{$R_{\rm lim}$}}
\newcommand{\tlim}{\mbox{$t_{\rm lim}$}}
\newcommand{\Mlim}{\mbox{$M_{\rm lim}$}}
\newcommand{\Mmin}{\mbox{$M_{\rm min}$}}
\newcommand{\Mmax}{\mbox{$M_{\rm max}$}}
\newcommand{\Mi}{\mbox{$M_i$}}
\newcommand{\Ncl}{\mbox{$N_{\rm obs}$}}
\newcommand{\Nobs}{\mbox{$N_{\rm obs}$}}
\newcommand{\tdis}{\mbox{$t_{\rm dis}$}}
\newcommand{\Mdis}{\mbox{$M_{\rm dis}$}}
\newcommand{\Mcross}{\mbox{$M_{\rm cross}$}}
\newcommand{\mcross}{\mbox{$M_{\rm cross}$}}
\newcommand{\tcross}{\mbox{$t_{\rm cross}$}}
\newcommand{\Halfa}{\mbox{H$\alpha$}}
\newcommand{\lm}{\mbox{$\log (\Mcl)$}}
\newcommand{\chinu}{\mbox{$\chi_{\nu}^2$}}
\newcommand{\tdisref}{\mbox{$t_4^{\rm dis}$}}
\newcommand{\Cref}{\mbox{$m_{\rm ref}$}}
\newcommand{\dd}{\mbox{{\rm d}}}

\title[Empirical determination of the disruption times of star clusters in four galaxies ]
{Star Cluster Formation and Disruption Time-Scales -- I.\\
An empirical determination of the disruption time of 
star clusters in four galaxies\thanks{
    Partly based on observations with the NASA/ESA Hubble Space Telescope,
    obtained at the Space Telescope Science Institute, which is
    operated by AURA, Inc., under NASA contract NAS 5-26555}
 }

\author[S. G. Boutloukos and H.J.G.L.M. Lamers]
       {
        S.G.Boutloukos$^{1,2}$
       \thanks{ Present adress: Theoretische Astrophysik, Universit\"at
                 T\"ubingen, Auf der Morgenstelle 10, D-72076,
                 T\"ubingen, Germany, email: 
                 {\tt stratos@tat.physik.uni-tuebingen.de}} 
          , and H.J.G.L.M. Lamers$^{1,3}$
       \thanks{email: {\tt lamers@astro.uu.nl}}
       \\
       $^1$ Astronomical Institute, Utrecht University, 
                 Princetonplein 5, NL-3584CC Utrecht, the Netherlands \\
       $^2$ Dept. of Physics, Section of Astrophysics, Aristotle University of
                 Thessaloniki, GR-54006, Thessaloniki, Greece\\
       $^3$ SRON Laboratory for Space Research, Sorbonnelaan 2,
                 NL-3584CC, Utrecht, the Netherlands
     }

\date{Received date/ Accepted date}

\pubyear{2002}

\begin{document}
\maketitle


\begin{abstract}
We derived the disruption times of star clusters from cluster samples of 
four galaxies: M51, M33, SMC and the solar neighbourhood.
If the disruption time of clusters in a galaxy depends only on
their initial mass as 
$\tdis ({\rm yrs}) = \tdisref \times (\Mcl/ 10^4 \Msun)^{\gamma}$, 
and if the cluster formation rate is constant,
then
the mass and age distributions  of the observed clusters,
will each be given by double powerlaw relations.
For clusters of low  mass or young age  the powerlaw depends on the
fading of the clusters below the detection limit
 due to the evolution of the stars. For clusters of high mass  
and old age the powerlaw depends on the disruption time of the clusters.
The samples of clusters
in M51 and M33, observed with $HST-WFPC2$ 
indeed show the predicted double powerlaw relations in both their
mass and age distributions. The 
values of $\tdisref$ and $\gamma$ can be derived from these relations. 
For the cluster samples
of the SMC and the solar neighbourhood, taken from the literature, 
only the age distribution is known. This also
shows the characteristic double powerlaw behaviour, which allows
the determination of $\tdisref$ and $\gamma$ in these galaxies.
The values of $\gamma$ are the same in the four galaxies within the
uncertainty
and the mean value is $\gamma= 0.62 \pm 0.06$. However the
disruption time $\tdisref$ of a cluster of $10^4$ \Msun\ 
is very different in the different galaxies. 
The clusters in the SMC have the longest disruption time,
$\tdisref \simeq 8 \times 10^9$ yrs, and the clusters at 
1 to 3 kpc from the
nucleus of M51 have the shortest disruption time of $\tdisref \simeq 4
\times 10^7$ yrs. The disruption time of clusters 1 to 5 kpc from the
nucleus of M33 is $\tdisref \simeq 1.3 \times 10^8$ yrs and for clusters 
within 1 kpc from the Sun we find $\tdisref \simeq 1.0 \times 10^9$ yrs.
\end{abstract}




\section{Introduction}
\label{sec:1}

{\bf In this paper we derive an {\it empirical relation between the
disruption time and the initial mass of clusters} 
in limited regions of the spiral galaxies M51 and M33. We
also derive the disruption times of clusters in the solar
neigbourhood of the Milky Way and in the Small Magellanic Cloud.
The determination of such an empirical 
 relation is important for several reasons:\\
(a) to compare the disruption of clusters in different
conditions of different galaxies,\\
(b) to explore and describe the evolution of the population of 
cluster systems in different galaxies using the simplest reasonable 
model,\\
(c) to predict and explain the evolution of field star
populations that result from cluster disruption.}

Stellar clusters in the Milky Way 
show a bimodal age distribution with peaks at the youngest age
(open clusters) and at the oldest age (globular clusters). This is 
because most clusters disrupt over time, due to the decrease
of mass by stellar evolution and due to tidal interactions. Only the most 
massive and most concentrated clusters, such as the globular clusters, 
survived after about 10 Gyrs.
This was first pointed out by Oort (1958), who noticed the lack of old
clusters with ages longer than about 1 Gyr in the Galaxy. He suggested
a time scale for the disruption of Galactic clusters of $5 \times
10^8$ yrs.

In fact, the dynamical evolution of clusters is determined by  
at least two timescales: \\
(a) the relaxation time $t_{\rm rxt}$, which is the time for
a  cluster to reach equilibrium between the kinetic energy
distribution of the stars and the potential well. Since the stars of a 
new-born cluster have a wide range in velocities and masses, the
interactions between the stars cause the massive ones to
accelerate the lower mass stars. As a result the low mass stars 
continuously gain energy and may leave the cluster whereas the massive
stars move deeper into the potential well. This effect includes
mass segregation and core collapse.\\
(b) the disruption time, \tdis, for interactions with
the surroundings. If the density of the cluster is not high
enough external factors like tidal forces and interactions with the
surrounding medium may cause a cluster to disrupt.
The disruption is accelerated by stellar evolution effects such as
mass loss by winds and by supernova explosions, which decrease the
density and the total amount of mass of the cluster.

Relaxation and disruption are two independent mechanisms which both 
occur in a cluster, albeit at different time scales, and both determine
the dynamical evolution of a cluster.

The first calculations of these timescales were made by Spitzer (1957).
He studied the dynamical effects of 
stellar clusters and derived an expression for the time needed 
for interstellar clouds to disrupt a cluster. He found a disruption
age of

\begin{equation} 
  \tdis=1.9 \times 10^8 \rho_{c}~~~~~{\rm yrs,}
\label{eq:tdisSpitzer}
\end{equation}
where $\rho_{c}$ is the mean density of a cluster in 
\Msunpc, for clusters with a density  $ 2.2 < \rho_c < 22$ \Msunpc. 
In a second paper Spitzer (1958) studied the relaxation time
of clusters. He found that  low mass clusters with a small radius 
have a short relaxation time, whereas the relaxation time is long
for massive and extended clusters. Therefore, 
massive and extended clusters may disrupt before they relax.
For a review see Wielen (1988).

Numerical simulations of the dynamical evolution of clusters have
been made by several groups, e.g. Chernoff \& Weinberg (1990),
de la Fuente Marcos (1997) and Portegies Zwart et al. (1998). 
Many of these studies concentrated on the dynamics of globular
clusters. For this study we are most interested in the evolution
of lower mass clusters.

Chernoff \& Weinberg (1990) made numerical simulations of 
clusters in the mass range of $3 \times 10^4 < \Mcl < 3 \times 10^5$
\Msun\ with three different density distributions and three different
stellar initial mass functions (IMF). They found that most of their
model clusters disrupt on a very long timescale  of $9 < \tdis < 40$ Gyrs.
Only the less massive, highly concentrated clusters with a steep
stellar IMF, survived disruption. 

The disruption of lower mass clusters with $100 < \Mcl < 750$ \Msun\
was studied by de la Fuente Marcos (1997), who used a single density
distribution for the numerical simulations and took the presence of
binaries into account. He found that all his model clusters disrupt.
The time at which this occurs increases with the cluster mass 
and depends on the adopted stellar IMF.

{\bf In this paper we derive the empirical relation between the
disruption time and the mass of clusters in different locations
of galaxies, based on the mass and/or age distributions of 
magnitude limited cluster samples. For M51 we use a cluster sample
in the inner spiral arms observed with $HST-WFPC2$ and studied by
Bik et al. (2002). For M33 we use the sample at the galactocentric
distance range of about 0.8 to 5 kpc, studied by Chandar et
al. (1999a, 1999b, 1999c) based on observations with $HST-WPC2$.
For the LMC we use the sample of clusters with ages derived by Hodge (1987)
based on photographic plates, and for the solar neighbourhood
we use the sample of clusters with ages derived by Wielen (1971). }

In Sect. 2 we describe the method to derive the disruption time
from the observed distribution of clusters over time and mass. 
In Sect. 3 we apply this method to the study of the clusters in 
M51 that were observed with the $HST-WFPC2$ camera. 
In Sect. 4 we derive the  disruption time of clusters in M33,
observed with $HST$ by Chandar et al. (1999a, 1999b, 1999c). In Sect. 5
we study the disruption time of clusters in the SMC,
observed by Hodge (1983, 1986, 1987) and in Sect. 6 we study
the disruption time of Galactic clusters within a distance of 1 kpc
from the Sun, from the sample studied by Wielen et al. (1971).
We compare the results of the different galaxies in Sect. 7.
Section 8 contains a critical discussion of our method and assumptions.  
The results are summarized in Sect. 9.


\section{The expected distribution of age and mass of clusters}
\label{sec:2}

In this paper we derive an empirical relation between the disruption
time and the mass of clusters, from the observed distribution of clusters
over age and mass. In this section we describe the concept and 
we discuss the expected age and mass distributions.

\subsection{The concept and the assumptions}
\label{sec20}

We assume a
constant cluster formation rate, i.e. the same number of clusters are
formed per unit time. We also assume a cluster initial mass
function (CIMF). Clusters are formed in the mass range of $M_{\rm min} < M <
M_{\rm max}$, with an assumed powerlaw probability $N(M) \sim
M^{-\alpha}$ with $\alpha>0$.
Because of this CIMF more clusters with low mass are formed than with
high mass.

Due to the evolution of the stars, clusters fade as they
age. 
\footnote{At specific wavelength bands, especially in the infrared,
clusters may actually show peaks in brightness at certain ages, e.g.
due to the formation of the first red supergiants at an age of
about 10 Myr, or the development of a asymptotic giant branch at
a later phase.}
Consequently at a certain age a cluster will fade below the detection
limit. For massive clusters this will occur at a higher 
age than for low mass clusters. This implies that there is a ``fading
limit'' in the mass versus age distribution of the clusters: for every 
cluster age there is a mass limit below which the clusters are too
faint to be detected. This limiting mass is higher for old clusters
than for young clusters.
The fading limit depends on the detection limit of the
instrument, the extinction and the distance of the parent galaxy of the cluster.
For sensitive instruments or nearby galaxies, the fading limit may be
at lower mass than the lower mass limit $M_{\rm min}$ of the clusters.

Suppose as a first order approximation that clusters keep their
initial mass until they are suddenly disrupted. We assume that 
clusters are disrupted at a certain age and that this
disruption age depends on their initial mass $M_i$ in such a way that massive
clusters survive longer than low mass clusters, for instance a
powerlaw of the type $\tdis(M_i) \sim M_i^{\gamma}$ with $\gamma>0$. 
This also implies the
existence of the inverse relation $M_{\rm dis}(t) \sim t^{1/\gamma}$.
Clusters with an initial mass $M_i$ which are older than 
their disruption time $\tdis (M_i)$  
will be disrupted. Similarly, clusters with an
age $t$ and with an initial  mass below $M_{\rm dis}(t)$ will also
be disrupted. We will show that the disruption relation $\tdis (M_i)$
can be derived from the mass and age histograms of clusters.

We summarize the assumptions:
\begin{enumerate}
\item{} the cluster formation rate is constant,
\item{} the clusters are formed with a fixed cluster initial mass
  function (CIMF),
\item{} {\bf the clusters have the same stellar initial mass function and
  the same lower mass cutoff,\footnote{The value of the adopted 
  lower mass cutoff does not affect
  the empirical determination of the dependence of \tdis\ on $M_i$
  directly. However, it may have affected the 
  derived masses of the clusters, if they are determined from integrated
  cluster photometry.}}
\item{} the clusters fade as they get older due to the evolution of their stars,
\item{} the clusters keep their initial mass untill they are suddenly
  disrupted,
\item{} the disruption time of a cluster depends on its inital mass.
\end{enumerate}

We realize that the assumption of sudden disruption is a severe
simplification of the real disruption. Theory and dynamical
models suggest that the mass of a cluster will decrease approximately
linearly with time (e.g. Spitzer 1957; Portegies Zwart et al. 1998). 
We have opted for this simplification for three reasons:\\
(i) it results in a very simple and easy to understand method to
derive the cluster disruption time from the mass and age
distributions of observed cluster samples.\\
(ii) the results can be used to quantitatively compare the cluster disruption
times in different galaxies or different locations in the same galaxy.\\ 
(iii) the disruption times derived from the age and mass
distributions of clusters with sudden disruption
are quite similar to those derived from the distributions of clusters 
with gradual disruption. This is shown in Appendix A.

{\bf We also realize that the disruption timescale of a cluster 
does not only depends on its mass but also on its density or radius
and on its environment. 
However, since we are interested in the dependence of $\tdis$ on mass,
for the reasons mentioned in the Introduction, we do not take the
density into account in this study\footnote{The empirical
determination of the density dependence of \tdis\ is the topic of a
study in progress.}. If clusters are approximately in pressure 
equilibrium with their environment, we can expect the density of all clusters
in a limited volume of a galaxy to be about the same, so the disruption
time of clusters in a limited volume of a galaxy will depend on the
mass (apart from clusters in highly excentric orbits). On the other
hand, if the initial density of clusters depends on their mass, e.g. as
$\rho \sim M^{x}$, and \tdis\ depends on mass and density
as $\tdis \sim M^a \rho^{b}$ then the empirical dependence of
\tdis\ on mass will be 
$\tdis \sim M^{\gamma}$ with $\gamma=a+bx$. If \tdis\ is proportional
 to the relaxation time $t_{\rm rxt} \sim M \rho^{-1/2}$ (Spitzer 1987)
we expect $a=1$ and $b=-1/2$, so $\gamma=1-x/2$.}


\subsection{The distribution in case of evolutionary fading only}
\label{sec:2a}

Even if there would be no disruption or evaporation of clusters, they
would still get fainter with time due to the evolution of their 
stars. This 
effect results in a decrease of the number of observable clusters as a
function of age for a given magnitude limit, described in this section.

Assume a constant cluster formation rate $S$ and a constant CIMF of
slope $-\alpha$. Then the number of clusters formed per unit time
and per unit mass in a certain area of a galaxy will be

\begin{equation}
\dd N(\Mcl,t)~=~S~\Mcl^{-\alpha}~\dd \Mcl~\dd t
\label{eq:CIMF}
\end{equation}
with $S$ in $\Msun^{\alpha -1}$~yr$^{-1}$,
where \Mcl\ is the {\it initial} stellar mass of the cluster.
This equation is valid for the mass range between the minimum and
maximum cluster mass. The constant $S$ is related to the 
cluster formation rate in the considered part of the galaxy.
Observations and theory show that $\alpha \simeq +2$ (Zhang \& Fall
1999; Whitmore et al. 1999; Bik et al. 2002). 
The total number of clusters formed per
unit time in the mass range of $M_{\rm min}$ to $M_{\rm max}$ is

\begin{equation}
\frac{ \dd N}{\dd t}~=~ \frac{S}{\alpha-1}~
(M_{\rm min}^{1-\alpha}-M_{\rm max}^{1-\alpha}) \simeq \frac{S}{M_{\rm min}}
\label{eq:Nformtot}
\end{equation}
if $\alpha=2$ and $M_{\rm max} >> M_{\rm min}$. The total mass of the 
stars formed per unit time in all clusters is

\begin{eqnarray}
\frac{ \dd M_{\rm tot }}{\dd t}~&=&~ \frac{S}{\alpha-2}~
(M_{\rm min}^{2-\alpha}-M_{\rm max}^{2-\alpha}) ~~~~{\rm if}~\alpha
\ne 2 \nonumber \\
 &=&~S~\ln(M_{\rm max}/ M_{\rm min})~~~~{\rm if}~\alpha=2
\label{eq:Mformtot}
\end{eqnarray}

Suppose that the brightness of a cluster in a particular wavelength
band, for instance in the $V$-band, decreases with 
time as a powerlaw, $F_V \sim t^{-\zeta}$ with $\zeta > 0$ 
for $t\simgreat10^7$ years, due to the evolution 
of the stars in the cluster (Leitherer et al. 1999, Bruzual \&
Charlot 1993). In that case 
the magnitude of a cluster varies with time and mass as

\begin{equation}
 V= \Cref -2.5 ~\log({\Mcl/10^4 \Msun})+2.5~\zeta ~
\log (t/10^8 {\rm yrs})
\label{eq:VMt}
\end{equation}
with $t$ in yrs and $\Mcl$ in \Msun. 
We have used the fact that for a given stellar IMF 
the flux of the cluster scales linearly
with the total mass of the cluster stars, \Mcl. 
The constant $\Cref$ depends on the distance and the mean extinction 
of the clusters and on the stellar
IMF and metalicity of the cluster stars.

\begin{equation}
\Cref =  M_{\rm V}^{\rm ref}-5+5 \times
\log{d({\rm pc})} +A_{\rm V}
\label{eq:cn}
\end{equation}
where $M_{\rm V}^{\rm ref}$
is the absolute visual magnitude of a cluster with an initial mass of
 $10^4 \Msun$ and an age of $10^8$ yr, which can be derived 
from cluster evolution models, and $A_{\rm V}$ is the mean extinction
in the observed region.
{\bf (For instance, $M_{\rm V}^{\rm ref}= -11.89$ magn. for Starburst99
 models of solar metallicity, a Salpeter IMF and a lower mass cutoff
 of 1 \Msun.)}
Let \Vlim\ be the limiting 
visual magnitude of the clusters that can be detected.
This implies that a cluster with a given mass \Mcl\ is detectable
at an age shorter than \tlim (\Mcl) with

\begin{equation}
\log \left(\frac{\tlim (\Mcl)}{10^8}\right)~=~\frac{-0.4 \times(\Cref - \Vlim)}{\zeta}
~+~ \frac{\log \Mcl/10^4}{\zeta}
\label{eq:tlim}
\end{equation}
Similarly, at an age $t$ only clusters with a mass larger than
\Mlim ($t$) can be detected, with

\begin{equation}
\log \left(\frac{\Mlim (t)}{10^4}\right) ~=~0.4 \times (\Cref - \Vlim)~+~\zeta \times \log (t/10^8)
\label{eq:Mlim}
\end{equation}

If there was no disruption of the clusters, the age distribution of the
observable clusters at any time would simply be 

\begin{eqnarray} 
\frac{\dd \Ncl}{\dd t} &=& \int_{\Mlim (t)}^{\Mcl(\rm max)} S~\Mcl^{-\alpha} ~d \Mcl
\nonumber \\
&\simeq& ~ 10^{-4-0.4(\Cref-\Vlim)(\alpha-1)}~\frac{S}{\alpha-1}~
\left(\frac{t}{10^8}\right)^{-\zeta(\alpha-1)}
\label{eq:Ntnodisruption}
\end{eqnarray}
for $t \simgreat 10^7$ years.
Here we have assumed that the maximum cluster mass, $\Mcl (\rm max)$,
 is much larger than the limiting mass at $t=10^7$ years, and we
assumed explicitely that the cluster formation rate $S$ is constant.
Similarly, the
mass  distribution of the observable clusters is

\begin{eqnarray}
\frac{\dd \Ncl}{\dd \Mcl} &=& \int_{0}^{\tlim (\Mcl)}  S~\Mcl^{-\alpha}~ d t 
\nonumber \\
&\simeq& ~ 10^{-4(\alpha-2)-0.4(\Cref-\Vlim)/\zeta}~S~
\left({\Mcl}{10^4}\right)^{-\alpha+ (1/\zeta)}
\label{eq:Nmnodisruption}
\end{eqnarray}
We see that under these assumptions both the mass distribution and the
age distribution of clusters are powerlaws of \Mcl\ and $t$
respectively. These powerlaws can be determined observationally.


\subsection{The distribution in case of cluster disruption}
\label{sec:2b}

Let us now assume that clusters disrupt at a time that depends 
only on their initial mass, as

\begin{equation}
\tdis (\Mcl)~ = ~ \tdisref ~ (\Mcl / 10^4 \Msun)^{+\gamma}
\label{eq:tdis}
\end{equation}
where $\tdisref$ is the disruption time (in yrs) of a 
cluster with an initial mass of $10^4$ \Msun.
The positive sign with $\gamma > 0$ 
indicates that the
disruption time is expected to be larger for the more massive clusters. 
Clusters with an age $t$ will only be observable if they have a mass
larger than $\Mdis=10^4(t/\tdisref)^{1/\gamma}$ and larger than the limiting mass
given by Eq. (\ref{eq:Mlim}). Alternatively, clusters with a mass
\Mcl\ will only be observable up to an age that is given by the
minimum of the ages in  Eqs. (\ref{eq:tlim}) and (\ref{eq:tdis}).  

If disruption is important, the observed age distribution of the clusters
will be

\begin{eqnarray} 
\frac{\dd \Ncl}{\dd t}& =& \int_{\Mdis (t)}^{\Mcl(\rm max)} S~\Mcl^{-\alpha}
d \Mcl \nonumber \\
  &\simeq&    \frac{S}{\alpha-1} 10^{-4(\alpha-1)}
\left(\frac{t}{\tdisref}\right)^{(1-\alpha)/\gamma}
\label{eq:Ntdisruption}
\end{eqnarray}
and the observed mass distribution will be 

\begin{eqnarray}
\frac{\dd \Ncl}{\dd \Mcl} &=&  \int_{0}^{\tdis (\Mcl)} S~ \Mcl^{-\alpha}~ d t ~
\nonumber \\
&\simeq& ~ S~\tdisref~10^{-4 \alpha}~\left(\Mcl/10^4 \right)^{\gamma-\alpha}
\label{eq:Nmdisruption}
\end{eqnarray}
We see that the slope of the mass distribution is less steep than that
of the CIMF, because disruption removes the low mass clusters first.

\begin{figure}
\centerline{\psfig{figure=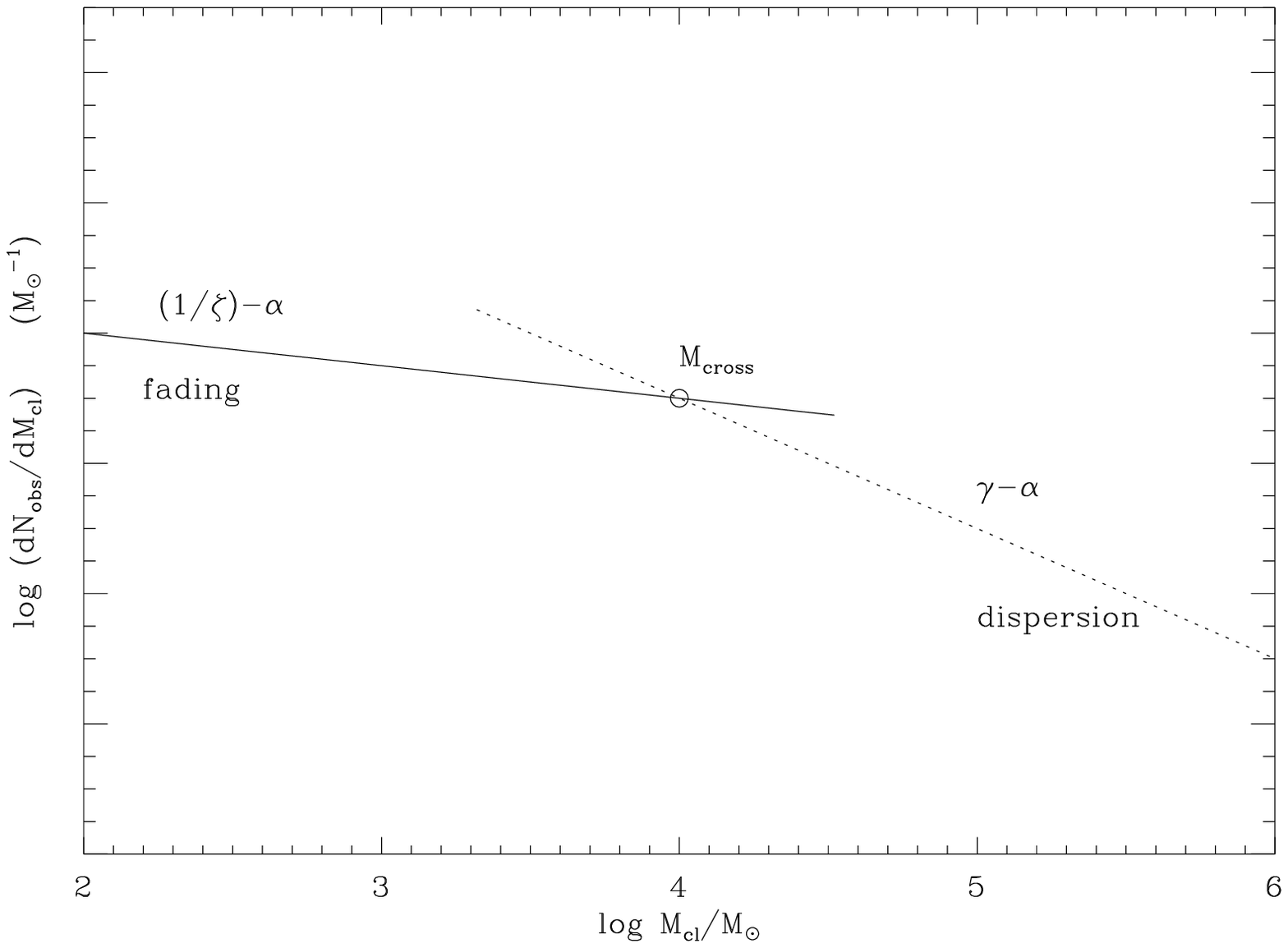,width=9.0cm}}
\centerline{\psfig{figure=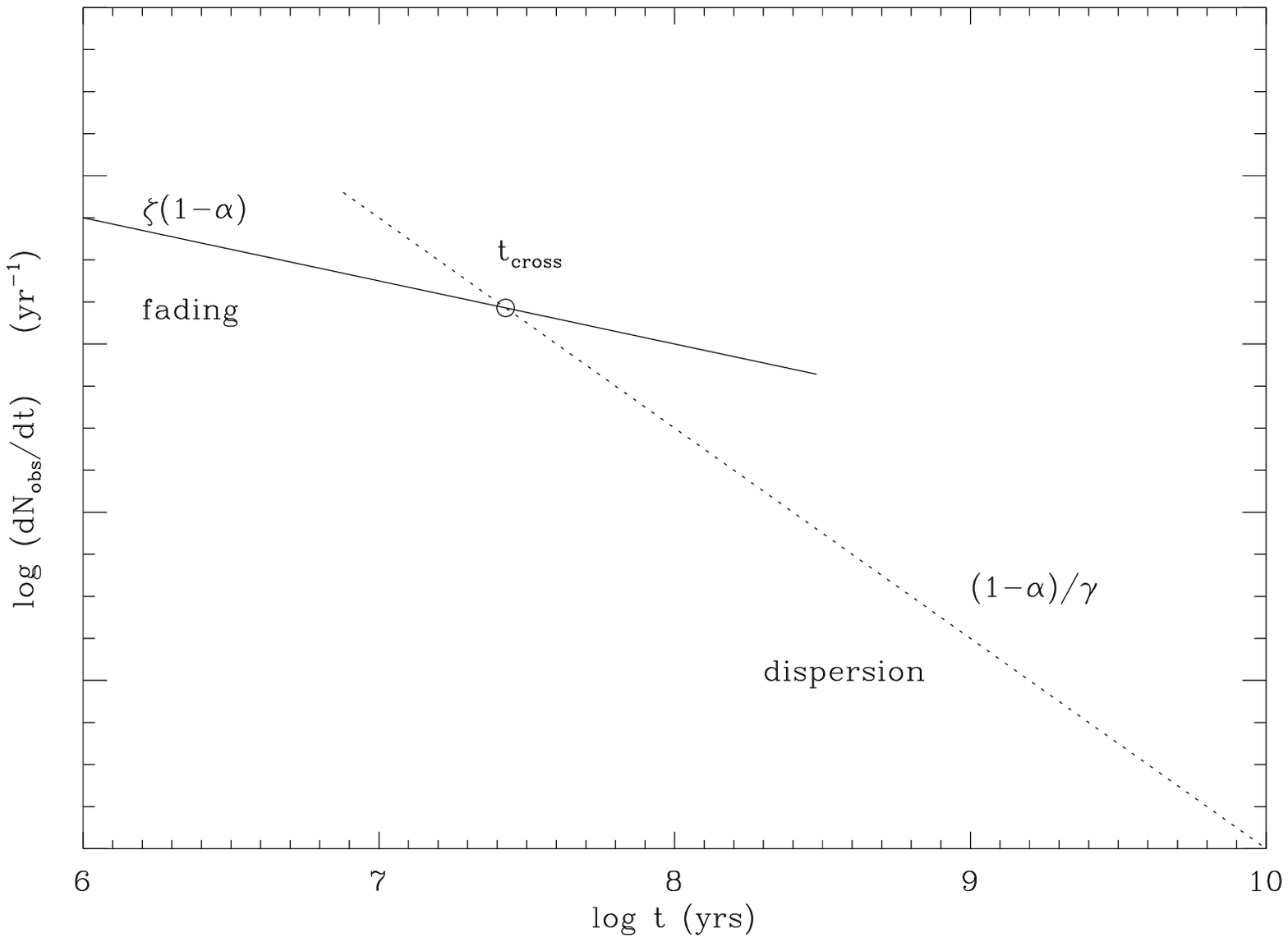,width=9.0cm}}
\caption{Schematic pictures of the 
predicted mass distribution (top figure) and age distribution 
(bottom figure) 
of clusters in case of no disruption and fading only (full lines) and
with disruption (dotted lines). The $y$-axes have arbitrary units.
The relation between the crossing points
$\Mcross$ and $\tcross$ is given by Eq. (\ref{eq:Mcrosstcross}). 
The slopes of the
different line sectors are indicated.} 
\label{fig:Npred}
\end{figure}

If both the disruption time and the observation limit are important,
the mass distribution and the age distribution of clusters will 
both consist of two powerlaws,
as shown in Fig. (\ref{fig:Npred}). 
This figure shows the expected relations  of $\log (d \Ncl /dt)$
as a function of $\log t$ and of $\log (d \Ncl /d\Mcl)$
as a function of $\log \Mcl$. 
The slopes are:

\begin{eqnarray}
\frac{1}{\zeta} - \alpha  ~~ &{\rm for}& ~~
 \log \left( \frac{\dd \Ncl}{\dd \Mcl} \right) = f( \log \Mcl) ~~~
{\rm fading} \nonumber \\   
\gamma - \alpha           ~~ &{\rm for}& ~~
 \log \left( \frac{\dd \Ncl}{\dd \Mcl} \right) = f( \log \Mcl) ~~~
{\rm disruption} \nonumber \\   
\zeta(1-\alpha)           ~~ &{\rm for}& ~~
 \log \left( \frac{\dd \Ncl}{\dd t   } \right) = f( \log t) ~~~
{\rm fading} \nonumber \\   
\frac{1-\alpha}{\gamma}   ~~ &{\rm for}& ~~
 \log \left( \frac{\dd \Ncl}{\dd t   } \right) = f( \log t) ~~~
{\rm disruption} 
\label{eq:slopes}
\end{eqnarray}   
So for a given value of the slope $\alpha$ of the CIMF, 
and a given value of the slope $\zeta$ (from cluster evolution models) 
we can derive the value of $\gamma$ from the observed slope of either 
the mass and or the age distribution of the clusters.
If both the mass and age distributions are known, the
values of both $\alpha$ and $\gamma$ can be derived.

 If fading line is not important (because the detection limit is so
good that all surviving clusters can be detected) then the fading part
of the age distribution will be horizontal. This follows from Eq.
(\ref{eq:Ntnodisruption}) with a lower limit of the integral of
$M_{\rm min}$.
If fading is not important and if
the disruption time $\tdis
 (\Mmin)$ for clusters with the mimimum mass would be longer than the
 age of the galaxy, the mass distribution would reflect the cluster IMF.  
Notice that the {\it shapes} of the mass and age distributions
in Fig. \ref{fig:Npred} are independent of the cluster formation 
rate and only depends on the
CIMF and on the disruption timescale. The {\it vertical shift} of these
distributions depend on the cluster formation rate.

The crossing points between the powerlaws in the two figures 
\ref{fig:Npred}
are indicated by \Mcross\ and \tcross. They give the cluster mass
and the cluster age at which disruption becomes important.
This depends on the value of $\tdisref$, i.e. on the constant of
the expression for the disruption time (Eq. \ref{eq:tdis}) as

\begin{equation}
\log (\tcross/10^8) ~=~ \frac{1}{1-\gamma \zeta}
\left\{ \log (\tdisref/10^8) + 0.4 \gamma (\Cref-\Vlim) \right\}
\label{eq:tcross}
\end{equation}
and

\begin{equation}
\log (\Mcross/10^4)~=~ \frac{1}{1-\gamma \zeta}
\left\{ \zeta \log (\tdisref/10^8) + 0.4 (\Cref-\Vlim) \right\}
\label{eq:Mcross}
\end{equation}
These two values of \Mcross\ and \tcross\ are related via Eq.
(\ref{eq:Mlim}) as

\begin{equation}
\log (\Mcross/10^4) ~=~0.4 \times (\Cref - \Vlim)~+~\zeta \times \log 
(\tcross/10^8)
\label{eq:Mcrosstcross}
\end{equation}
This implies that we can derive the value of $0.4\times(\Cref -
\Vlim)$ 
by comparing the empirical values \Mcross\ and \tcross, and then we can
derive the value of $\tdisref$ from \Mcross\ or \tcross\
 (Eqs. (\ref{eq:Mcross}) and (\ref{eq:tcross})).

We will show below that the observed mass and age distributions of
cluster samples in the different galaxies indeed show two powerlaw slopes.
>From the empirical slopes we can test the assumptions of this simple
model and derive the dependence of the disruption time on
the cluster mass, given by $\tdisref$ and $\gamma$ (Eq. \ref{eq:tdis})
\footnote{We have derived the expressions of the mass and age
  distributions under the assumption that the cluster sample is
magnitude limited in the $V$-band. The expressions can easily be 
modified for a sample that is magnitude limited in any other band. 
In that case only the values of $M_V^{\rm ref}$, $\zeta$ and $\Cref$ 
have to de adapted.}.

In this section we have assumed that the cluster disruption occurs
instantaneously when the cluster reaches an age $\tdis (\Mcl)$.
That means, a cluster keeps its mass (apart form stellar evolution
effects) until it suddenly disrupts.
This is a severe assumption, that allows us to determine 
from the observed mass and age distributions how the
disruption time of clusters depends on their initial mass.
In Appendix A we show that the mass and age distributions
of gadually disrupting clusters are very similar to those of
instantaneously disrupting clusters, if the expressions for the
disruption time are the same in both cases.

\section{A study of the disruption of clusters in M51}
\label{sec:3}

M51 or the Whirlpool galaxy (NGC 5194) is a grand design Sc spiral galaxy
that interacts with the dwarf galaxy NGC 5195. Their
close encounter is estimated to have occurred about $250-400$ Myr
ago  (Hernquist 1990). Salo \& Laurikainen (2000) argued that there
may even have been a more recent encounter about 50 to 100 Myrs ago.
 The whole
system lies in a distance of approximately 8.4 Mpc (Feldmeier et
al. 1997).
The Whirlpool galaxy has an almost face-on orientation,
with an inclination angle of 23 to 35 degrees (Monnet et al. 1981).
This makes it ideal for the study of its 
cluster formation history.

Scuderi et al. (2002) found that the interaction with its 
companion produced a huge starburst in the nucleus. 
Lamers et al. (2002) suggested that the formation of 
primarily massive stars in
the bulge of M51 is also due to the interaction. 
Bik et al. (2002), hereafter referred to as Bik02, 
studied clusters in the inner spiral arms
and derived the initial mass function of the clusters, as well as the
cluster formation rate. They found no evidence for an increased
cluster formation rate due to the interaction in this region.
The ages and masses of the M51 clusters
derived by Bik02 will be adopted
in this study.


\subsection{The observations}
\label{sec:3a}

\begin{figure}
\centerline{\psfig{figure=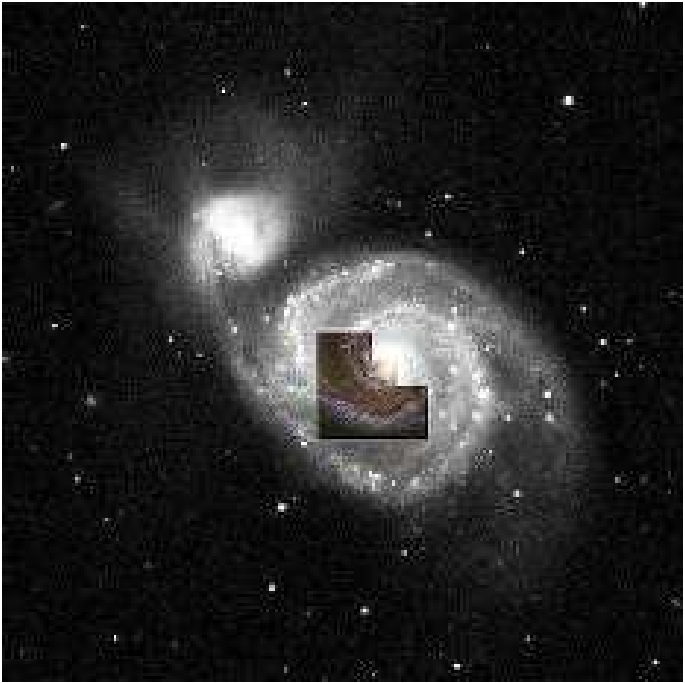, width=9.0cm}}
\caption{The area of M51 observed with the $HST-WFPC2$ camera, 
superimoposed on the image of M51.}
\label{fig:locM51}
\end{figure}

\begin{figure}
\centerline{\psfig{figure=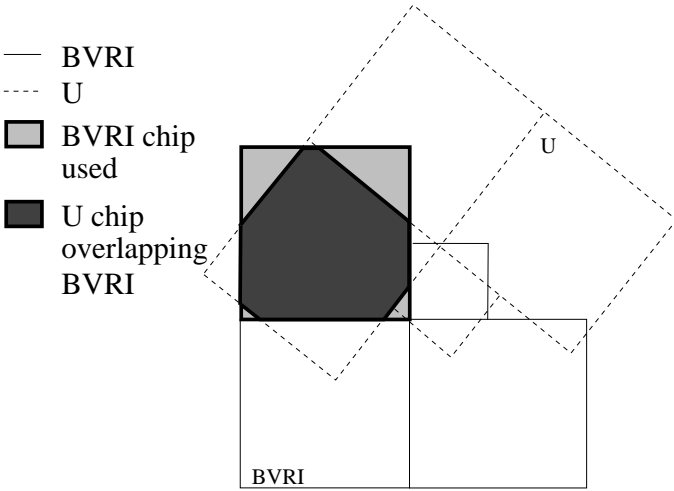, width=9.0cm}}
\caption{The orientation of the $WFPC2$ camera at different 
observing epochs. The grey-scaled area is the part of the image
in which the clusters, studied by Bik et al. (2002) and used in this
paper, are located.
The center of this area is at RA(2000)=$13^h29^m56.^s9$
and DEC(2000)=$47^012'22".1$.
The dark region is the area where the $U$-image overlaps 
the $BVRI$-images.  Each $WF2$-chip covers an area of $3.25\times3.25$
kpc.   The nucleus of M51 is close to the center
of the $PC$-image (i.e. the small rectangular area in the center).}
\label{fig:locWFC}
\end{figure}

 The M51 system was observed with the $HST-WFPC2$ in five wideband
 filters ($U$, $B$, $V$, $R$ and $I$)
 as part of the HST Supernova Intensive Study program. The
 observations have been described by Scuderi et al. (2002) and Bik02. 
Bik02 detected 877 clusters in an area of about 
$3.2 \times 3.2$ kpc about 0.8 to 3.1 kpc from the nucleus, 
observed with the $WF2$
camera. The area studied by Bik02 is shown in Figs. (\ref{fig:locM51})
and  (\ref{fig:locWFC}).
 All clusters were
detected in at least three bands, including $V$ and $R$.
If a cluster is detected in only three or four bands, the empirically
 determined lower magnitude limits were adopted for the other bands.

Bik02 determined the ages, the initial masses 
and the values of $E(B-V)$ of
these clusters using the 3/2DEF-method, which is a least-square fit
comparison
between the observed  energy distribution and the predicted energy 
distributions of clusters of different extinction, initial mass and 
age from  the Starburst99 models of Leitherer et al. (1999).
The cluster models for instantaneous starformation,
with solar metallicity and with a  
stellar IMF of $dN/dM \sim M^{-\alpha}$ with $\alpha=2.35$
were adopted. Bik02 showed that the models with solar
metallicity fit the observed energy distributions better than those
with twice solar metallicity. They also found from the lack of
[OIII] line-emission, in comparison with the strength of the \Halfa\ 
emission, that
the stellar mass function of the clusters is truncated above about
30 \Msun. So they adopted the cluster parameters derived from the
models with solar metallicity and with an upperlimit for the stellar
mass of 30 \Msun\ and a lower limit of 1 \Msun.
These models were calculated using the Starburst99
program by N. Bastian (private communication). 
The Starburst99 models do not extend beyond an age of 1 Gyr.
For older clusters, in the range of 1 to 10 Gyr,  
Bik02 adopted the energy distributions  for solar metallicity clusters
with a stellar IMF of $\alpha=2.35$ 
and an upper mass limit of 25 \Msun, using the
Frascati evolutionary tracks (Romaniello, private communication).
 
The least square fit of the model energy distributions to
the observed ones resulted in estimates of the
mass, age and $E(B-V)$ of the clusters and their uncertainties. 
The quality of the fit is given by
the value of the reduced $\chi^2$ ($\chinu$) for each cluster.
(For the details of these determinations and their 
accuracy, the reader is referred to Bik02.)
We will use the ages and the initial masses of the
clusters determined by Bik02 to derive 
the disruption time of clusters in
M51, with the method explained in Sect. \ref{sec:2}.

We point out that the cluster mass, determined in this way 
is the total {\it initial} mass of the stars that are still in the
cluster. This means that it is corrected for the stellar evolution
effects. However it is not corrected for stars that may have left the
cluster. Stars that have left the cluster and are at a distant of more
than about 0.4 arcsec, corresponding to about 16 pc, do not contribute
to the photometry and their mass is not counted. 
(A radius of 4 pixels or 0.4 arcsec 
was used by Bik02 for measuring the photometry
of the clusters.) 
{\bf The relatively large radius of 16 pc used for measuring the
  magnitudes implies that our sample is basically magnitude limited,
rather than surface brightness limited as the vast majority of the
clusters is expected to have a radius less than 16 pc. For instance,
Chandar et al. 1999b found that the bright clusters in M33, which is
about ten times closer than M51, have a core radius of only 0.1 to a
few pc and a FWHM radius about twice as large.}


\subsection{The detection limit}
\label{sec:3b}

\begin{figure}
\centerline{\psfig{figure=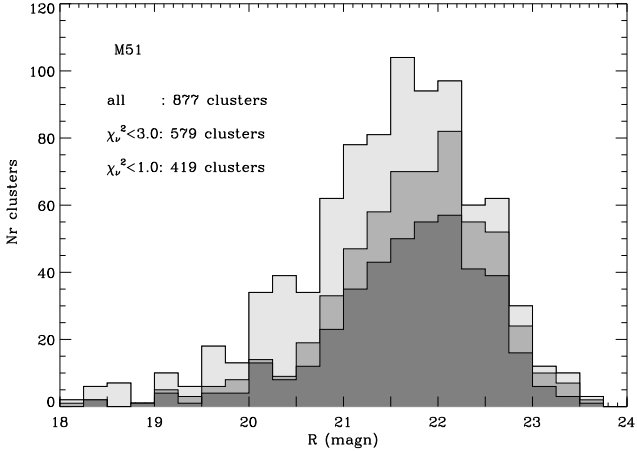,width=9.0cm}}
   \caption{The histogram of the $R$-magnitudes of the M51 clusters 
studied by Bik et al. (2002). The lightest distribution is for all
877 clusters measured. The other two distributions are for subsets
of clusters whose energy distibution could be fitted to cluster models
with an accuracy of $\chinu < 1.0$ and 3.0.
The turnover is at $R \simeq 22.0$.
We adopt this value as the magnitude limit of our sample.}
   \label{fig:histobsmag}
 \end{figure}

The observational detection limit of clusters in M51 depends 
on the accuracy of the observations, in this case the $HST-WFP2$ 
photometry.
For the determination of the detection limit, we will use the 
$R$-magnitude, because it is less sensitive to extinction than the
photometry in the $V$ band, and because all clusters were detected in the 
$R$-band.
 Figure (\ref{fig:histobsmag}) shows the histogram of the 
$R$-magnitude
of the sample of 877 clusters studied by Bik02. 
The figure shows three samples: the full sample, and the two samples of
clusters for which the observed energy distribution could be fitted
with a model with an accuracy of $\chinu \le 3.0$ and 1.0 respectively.
For all three samples the 
number of clusters increases towards fainter clusters 
from $R \simeq 18$ to $R \simeq 22$, as expected for a 
cluster IMF that predicts many more low mass clusters than high mass
ones. For all three samples the turn-over occurs near $R \simeq 22$ 
and the distribution drops steeply to fainter clusters. The
magnitude halfway down this steep drop is $R \simeq 22.5$, and the faintest
clusters have $R \simeq 23.7$. We will adopt the conservative 
upperlimit of $R = 22.0$ for the completeness limit of our sample.

The detection limit in $R$-magnitude corresponds to a certain
lower limit in mass and age of the clusters. This relation
can be derived from the Starburst99 cluster models of Leitherer et al. (1999).
These models show that a cluster with instantaneous starformation at
solar metallicity and with a stellar IMF of $\alpha=2.35$ for 
stars in the range of 
$1 < M_* < 30$ \Msun, with a total mass of $10^4$ \Msun\
and an age of $10^8$ yrs has an $HST$ $R$-band magnitude of 
$M_R=-8.37$. This implies that $m_{\rm ref}^{R}=21.61$ 
for a distance of 8.4 Mpc and a mean extinction of $A_R=0.36$
(Bik02).

\begin{figure}
\centerline{\psfig{figure=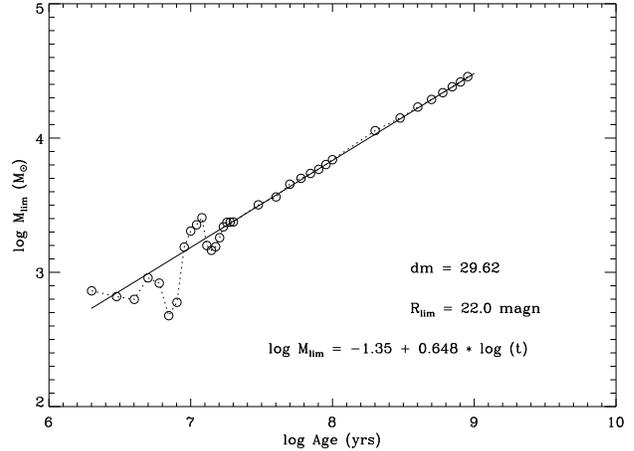,width=9.0cm}}
\caption{The relation between the limiting 
age and initial mass of instantaneously
formed clusters of solar metallicity with $\alpha=2.35$
(from Leitherer et al. 1999) at a distance of 8.4 Mpc 
with a mean extinction of $A_R=0.36$ and for a given 
observational magnitude limit of $\Rlim = 22.0$. 
The straight full line gives a powerlaw fit to the data for 
$\log t \ge 7.3$ with a slope of $d ~\log (M_{\rm lim})/d~ \log(t)=0.648$.
}
\label{fig:detectionlimit}
\end{figure}

The detection limit as a function of time 
is shown in Fig. (\ref{fig:detectionlimit}).
The figure shows a perfect linear relation 
for clusters with age $\log t \ge 7.3$, which is given by

\begin{equation}
\log \Mlim~\simeq~-1.350~+~0.648 \times \log t (\rm yrs)
\label{eq:limtheory}
\end{equation}
For smaller ages the relation is irregular with dips at
$\log t \simeq 6.9$ and $\log t \simeq 7.2$, due to the 
presence of red supergiants.
In this study we will use the exact dependence of \Mlim\ on age,
shown by the dashed line in Fig. (\ref{fig:detectionlimit}). The
linear approximation is only used for checking the results and for
easy comparison with the predicted age and mass
distribution, derived in Sect. \ref{sec:2}. 
Comparing Eq. (\ref{eq:limtheory}) 
with Eq. (\ref{eq:Mlim})
we see that $\zeta = 0.648$ and $0.4 \times(m_{\rm ref}^{R}-\Rlim)=-0.16$
for $\log \Mcl \simgreat 3.4$ and $\log t \simgreat 7.3$.

\subsection{The clusters}
\label{sec:3c}

\begin{figure}
\centerline{\psfig{figure=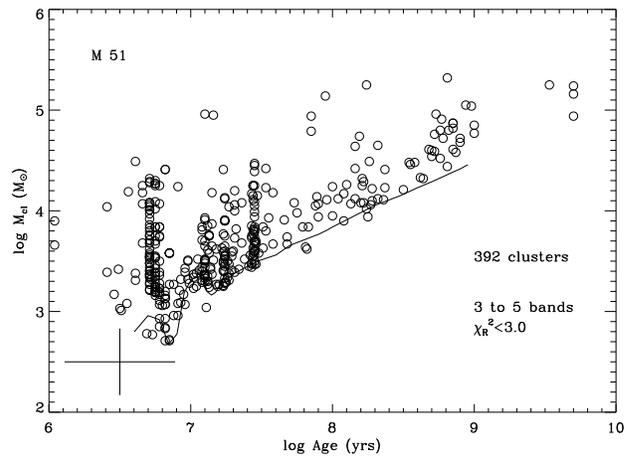, width=9.0cm}}
\caption{The relation between the initial mass \Mcl\ 
(in $M_{\sun}$) and age (in yrs) of 392 M51 clusters 
with $M_V<-7.5$, whose energy
distribution could be fitted to a cluster model with an
accuracy of $\chi_{\nu}^2 < 3.0$.
The full 
line is the detection limit for a limiting magnitude of $\Rlim=22.0$
at a distance of 8.4 Mpc without extinction. The cross indicates 
the size of typical errorbars (see Bik et al. 2002).
}
\label{fig:M51mt}
 \end{figure}

The initial mass versus age distribution of the clusters studied by
Bik02 is shown in Fig. (\ref{fig:M51mt}). We only show this
distribution for clusters with $M_V < -7.5$, 
for which the fit of the energy distribution has
$\chinu \le 3.0$. The distributions of the full sample and of the
sample of clusters with $\chinu \le 1.0$ show the same characteristics
as the one shown here.
%
%
The distribution shows several features:\\
(a) The lower limit of the data, agrees with the predicted limit
for instantaneously formed clusters and for a detection limit of 
$\Rlim=22.0$, derived above. Even the predicted dips 
near $\log t = 6.9$ and 7.2 may be present.\\
(b) The clusters older than about $10^8$ years all have initial masses 
in excess of $\log (\Mcl)\simeq 3.8$ whereas the younger clusters have masses
down to about 500 \Msun.\\
(c) There are concentrations in the distribution at $\log(t)=6.70$ and
7.45 and possibly also near 7.2. These are due to statistical effects
in the age range of $6.5 < \log(t) < 7.5$, where the colours of the
cluster models change rapidly with time (see Bik02). \\
(d) There is a drop in the density of the points at $\log (t)>
7.5$. Since the age scale is logarithmic, we might have expected an
increasing density towards higher age if the cluster formation
rate was constant and if all formed clusters would have survived
disruption. \\
(e) The presence of massive clusters with a age of about 5 Gyr
shows that either the disruption time in M51 is very long, 
or that massive clusters disrupt much slower than low mass
clusters. We will show that the first possibility disagrees
with the observed mass and age distributions.\\
(f) The general shape of the observed mass versus age 
distribution agrees with that expected for gradual disruption
with a mass dependent disruption time (See Appendix 1).

Bik02 determined the cluster initial mass function (CIMF)
from the mass distribution of clusters younger than 10 Myr.
These clusters are not yet affected by disruption. They
derived a powerlaw with a slope of 

\begin{equation}
\dd \log (\Nobs) / \dd \log (\Mcl)~=~- 2.1 \pm 0.3
\label{eq:cimf}
\end{equation}
in the mass range of $3.4 < \log(\Mcl) < 5.0$.
Bik02 also studied the cluster formation history from the age
distribution of the observed clusters. They used clusters
with an initial mass in excess of $10^4~\Msun$. They found that the
formation rate of these clusters has been increasing continuously with
time from about 10 Gyr up to the present. They argue that 
this {\it apparent} increase in the cluster formation rate over such a 
long time is due to the disappearance of clusters with increasing age 
due to disruption.
So we assume for this study that
the real cluster formation rate in M51 has been about constant.
We return to this assumption in the discussion.


\subsection{The determination of the cluster disruption time}
\label{sec:3d}

\begin{figure}
\centerline{\psfig{figure=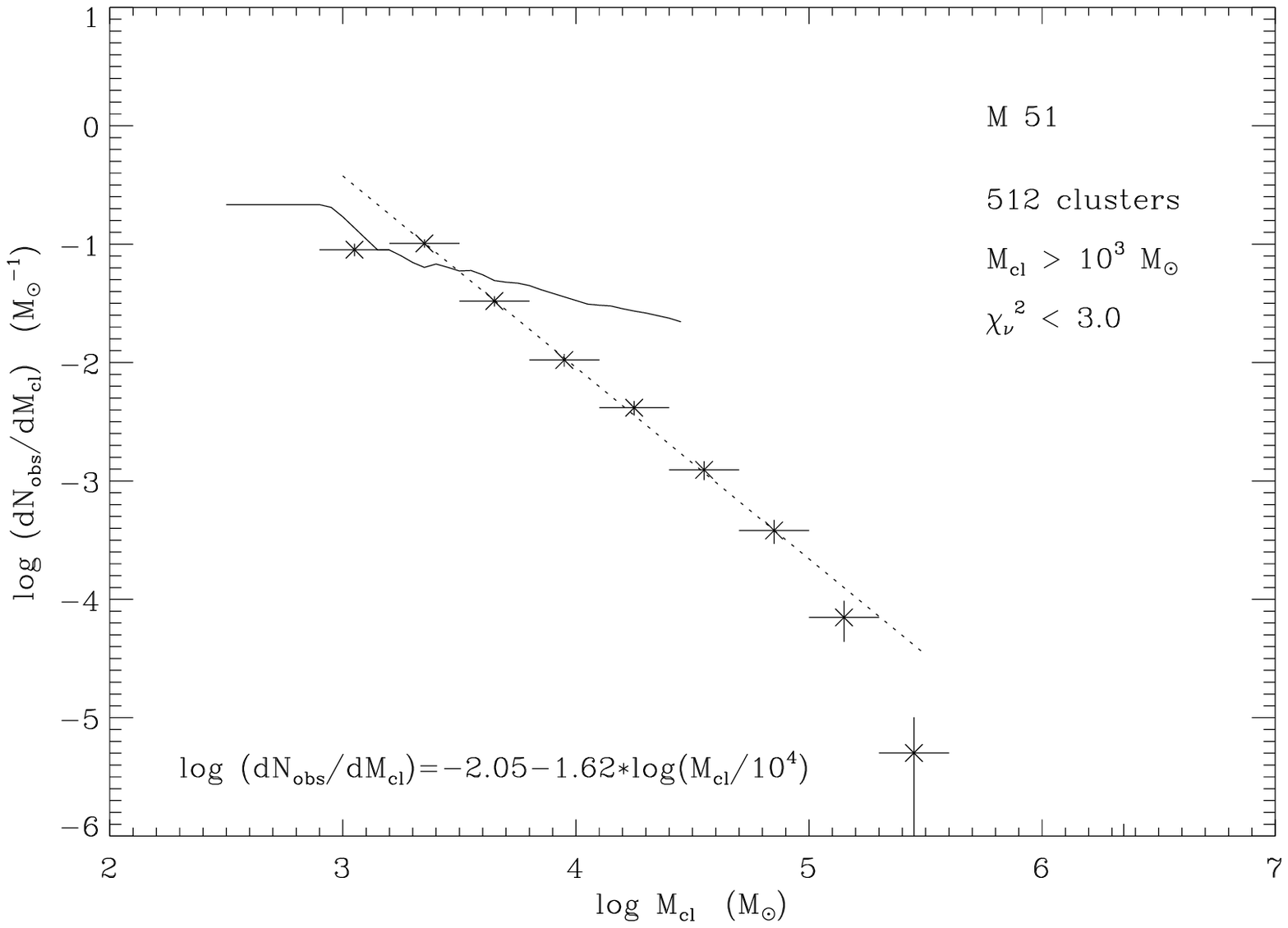,width=9.0cm}}
\centerline{\psfig{figure=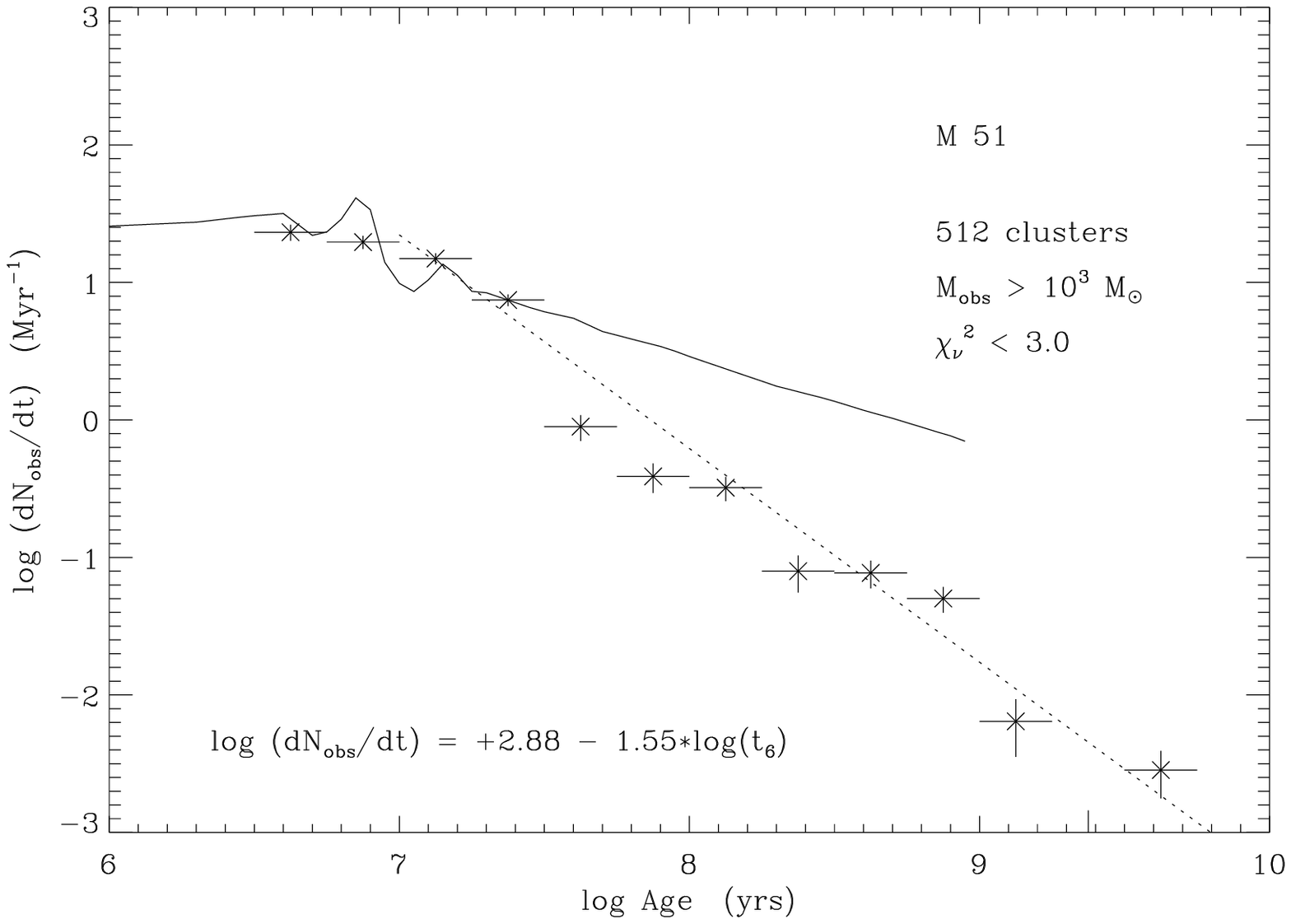,width=9.0cm}}
\caption{The mass (upper=Fig a.) and age (lower=Fig b.) 
distribution of the 512 observed clusters in M51 with $\Mcl > 10^3~\Msun$
and $\chinu \le 3.0$. The distribibutions are
fitted with two lines: the full lines show the expected decrease
due to evolutionary fading below the detection limit in case of no disruption. 
The dashed lines are the least square
powerlaw fits of the mass distribution of 
clusters with $3.3 \le \log( \Mcl) \le 5.2$ (Fig
a.) and the age distribution of clusters with $7.0 \le \log (t) \le 9.8$
(Fig b.), i.e. with disruption. The least square fits are given in the figure.
 From these lines the
dependence of the disruption time on initial mass can be derived.}
\label{fig:M51dist}
\end{figure}

The mass distribution of the 512 clusters with an accurate mass
determination, $\chinu < 3.0$, is shown in Fig. (\ref{fig:M51dist}a)
and the age distribution of these clusters 
 is shown in Fig. (\ref{fig:M51dist}b).
In this sample we only include clusters with an initial mass
in excess of $10^3$ \Msun, in order to avoid the sample
being ``polluted'' by the brightest stars.
The vertical axes give respectively $d \Nobs / d \Mcl$ in
number per \Msun\ and $\dd \Nobs / \dd t$ in number per Myr. 
The vertical error bars indicate the $1
\sigma$ Poisson errors. The horizontal error bars indicate the width of
the adopted intervals. Both  distributions clearly show
a trend of decreasing number with increasing mass or age.
The full sample of all 877 clusters, not shown here, has 
the same distribution, but with a larger scatter, due to the larger
uncertainties in the ages and masses.

In Sect. \ref{sec:2} we have predicted that if evolutionary 
fading and disruption both result in a limiting detectable 
mass that depends as a powerlaw on the age,
then both the mass and the age distributions of the
observed clusters should show a double powerlaw relation,
 given by  Eqs. (\ref{eq:Ntnodisruption}), 
(\ref{eq:Nmnodisruption}), (\ref{eq:Ntdisruption}) and
(\ref{eq:Nmdisruption}).  
Here we adopt the $M(t)$ relation of the detection limit
as shown in Fig. (\ref{fig:detectionlimit}) in  case of
no disruption and the relation Eq. (\ref{eq:tdis}) for higher ages and
masses with disruption.

\subsubsection{Fitting the mass and age distributions}
\label{Sec:3d1}

The mass distribution of the clusters, shown in Fig.
(\ref{fig:M51dist}a), decreases to higher masses much faster than
expected for evolutionary fading only. 
We have 
numerically  calculated the expected mass distribution in case of no
disruption, using the first part 
of Eq. (\ref{eq:Nmnodisruption}) for a CIMF of $\alpha=2$. 
This is
the value that was found by Zang \& Fall (1999) and Whitmore et al. (1999)
for the Antennae galaxies and it agrees with the value of $2.1 \pm
0.3$ derived for M51 by Bik02.

The predicted relation for
evolutionary fading (full line) has been vertically adjusted to fit
the data points in the smallest mass bins. This shift 
 is not well determined, because it depends on only two points. This
 implies that the crossing point of the two lines 
for evolutionary fading and for disruption of the mass distribution 
is not well determined. 

We have fitted a powerlaw of the type

\begin{equation}    
\log(\dd \Nobs /\dd \Mcl) = a_0 + a_1 \times \log (\Mcl)
\label{eq:dndmpowerlaw}
\end{equation}
to
the data  in the mass range of $3.5 \le \log (\Mcl) \le5.25$ 
by means of a weighted linear regression. The data point at
\lm = 5.45 was not taken into account because this appears to be close
to the mass upperlimit. The powerlaw has a slope in
of $-1.62 \pm 0.09$. 
We have done the same analysis for the full cluster sample 
and for the sample of clusters with $\chinu < 1.0$. The resulting 
parameters of the fits are listed in Table (\ref{tbl:M51fits}). We see that
the slopes of the powerlaw fits are very similar for the three
samples, but that
the value of the constant $a_0$ scales approximately with 
the log of the number of clusters in the three samples, as expected.
Taking the weighted mean of the three determinations, we find 
a mean value of $<a_1>=-1.54 \pm 0.07 $.

The age distribution is shown in Fig. (\ref{fig:M51dist}b).
The predicted relation due to fading (full line) has been fitted
through the four youngest age bins. The observed distribution
for $\log (t) > 7.5$ is far below the fading line and decreases
steeper with age than the fading line.
We have fitted a powerlaw expression of the type

\begin{equation}
\log (\dd \Nobs /\dd t)    = b_0 + b_1 \times \log (t)
\label{eq:dndtpowerlaw}
\end{equation}
to the data. The relation was forced to
extend to the last bins on the fading line, i.e. at $\log(t) = 7.1$ and
7.4. The parameters of the powerlaw fit are given in Table 
(\ref{tbl:M51fits}), for the sample shown here as well as for the other two
cluster samples. We find that
$<b_1>=-1.49 \pm 0.07$.

The points where the linear extensions of the fading lines and the powerlaw
fits in the mass and age distributions of Fig. (\ref{fig:M51dist}) 
cross is
\begin{eqnarray}
\log(\Mcross)~&=&~3.4 \pm 0.5~~~~~{\rm in~} \Msun \nonumber \\
\log(\tcross)~&=&~7.3 \pm 0.3~~~~~{\rm in~ yrs} 
\label{eq:M51cross}
\end{eqnarray}

The empirical values of $<a_1>$ and $<b_1>$, 
together will the values of \Mcross\ and \tcross,
 will be used to derive the cluster disruption time
in Sect. \ref{sec:3d3}.

\begin{table*}
\caption[]{Powerlaw fits to the mass and age distributions of 
M51 clusters\\
$\log(\dd \Nobs /\dd \Mcl) = a_0 + a_1 \times \log (\Mcl/10^4)$~~~ in $\Msun^{-1}$ \\
$\log(\dd \Nobs /\dd t)~~~~ = b_0 + b_1 \times \log (t/10^8)$~~~~~~~ in
${\rm yr}^{-1}$ }
\begin{tabular}{ll|ccc|ccc}
\hline\hline\noalign{\smallskip}
Sample & Nr & $a_0$ & $a_1$ & $\log (\Mcross)$ & $b_0$ & $b_1$ & $\log
(\tcross)$ \\
\hline
\noalign{\smallskip}
All              & 744 & -1.87 & -1.48 $\pm$ 0.06 & 3.4$\pm$0.5 & -6.04 & -1.43 
$\pm$ 0.05 & 7.3$\pm$0.3 \\
$\chinu \le 3.0$ & 512 & -2.05 & -1.62 $\pm$ 0.09 & 3.4$\pm$0.5 & -6.22 & -1.55
$\pm$ 0.07 & 7.2$\pm$0.3 \\
$\chinu \le 1.0$ & 380 & -2.18 & -1.61 $\pm$ 0.11 & 3.4$\pm$0.5 & -6.34 & -1.57
$\pm$ 0.09 & 7.2$\pm$0.3 \\
\hline 
\end{tabular}\\
\label{tbl:M51fits}
\end{table*}

\subsubsection{The cluster formation rate}
\label{sec:3d2}

The predicted distributions of the fading lines were
fitted through the first two mass bins and the first four age bins
in Fig. (\ref{fig:M51dist}).
This vertical adjustment depends on the detection limit and on the
cluster formation rate in the observed region of M51, as given by
Eqs. (\ref{eq:Nmnodisruption}) and (\ref{eq:Ntnodisruption}). 
Adopting the value of $\zeta=0.648$ and $0.4 \times (\Cref-R_{\rm
  lim})=-0.16$ (see Sect. \ref{sec:3b}) we find that 
$\log~ S= -1.70 \pm 0.20$ and $-1.66 \pm 0.10$ 
(in Nr~\Msun/Myr) from the mass and the age
distribution respectively. 
The value derived from the mass
distribution is more uncertain than that from the age distribution,
because only two data-bins were used for fitting the fading line to
the mass distribution. Nevertheless, we see that the value of $S$
derived in two independent ways is about the same.
The cluster formation rate  in the
observed region of M51 is 
about $10^{-1.66}~\Mcl^{-2}$ in Nr~\Mcl$^{-1}$~Myr$^{-1}$ 
(see Eq. (\ref{eq:CIMF})). 
We conclude that the mean 
cluster formation rate in the observed area of M51 is
(see Eq. (\ref{eq:Nformtot}) with $M_{\rm min}=10^3~\Msun$ )

\begin{equation}
\log (\dd \Ncl/\dd t)_{\rm form}=~ -4.7 \pm 0.10 ~~~~~ {\rm in~Nr~clusters~yr}^{-1} 
\label{eq:M51formationrate}
\end{equation}
This is for clusters with $\Mcl > 10^3~\Msun$. The mass of the stars
formed in the clusters per unit time is (see Eq. (\ref{eq:Mformtot}))

\begin{equation}
\log (\dd M_{\rm tot}/\dd t)_{\rm form} 
~=~ -0.9 \pm 0.10 ~~~~~ {\rm in}~ \Msun~{\rm yr}^{-1}  
\label{eq:M51formationrate}
\end{equation}
where we adopted $M_{\rm min}=10^3$ and $M_{\rm max}=3~10^5$
\Msun. This is the cluster formation rate in an area of 
3.25$\times$ 3.25 kpc about 1 to 3 kpc from the nucleus (see Fig. 
(\ref{fig:locWFC})).


\subsubsection{The cluster disruption time}
\label{sec:3d3}

We compare the empirical slope of the powerlaws  of  
$<a_1>= -1.54 \pm 0.07 $ and
$<b_1>= -1.49 \pm 0.07$, derived from the
distributions in Fig. (\ref{fig:M51dist}), with the predicted
slopes of $\gamma - \alpha$ and $(1 - \alpha)/\gamma$
respectively (Eqs. (\ref{eq:Nmdisruption}) and
(\ref{eq:Ntdisruption})). This gives the following two estimates of
$\gamma = 0.46 \pm 0.07$ and $\gamma = 0.67 \pm 0.03$ if $\alpha=2.0$.
We see that the two determinations of $\gamma$ 
from the mass and age distributions differ significantly.
We adopt the mean value of $\gamma \simeq 0.57 \pm 0.10$.

The constant $\tdisref$ in the expression for the disruption time
(Eq. (\ref{eq:tdis})) can be derived from the values of 
\tcross\ and \mcross\ where the fading and the disruption lines
in Fig. (\ref{fig:M51dist}) cross. Their values are given in Eq.
(\ref{eq:M51cross}). These values were derived independently from one another
from the empirical age and mass distributions of
Fig. (\ref{fig:M51dist}). We have predicted that the values of \tcross\
and \Mcross\ should be related to one another by means of the relation of
Eq. (\ref{eq:Mcrosstcross}). From the values of $\log \tcross = 7.3 \pm
0.3$, $0.4(\Cref-R_{\rm lim})=-0.16$ (Sect. \ref{sec:3b}) and $\zeta=0.648$
we predict that $\log \Mcross=3.38 \pm 0.19$. The observed value of
$3.4 \pm 0.5$ agrees excellently. So the values of \Mcross\ and
\tcross\ are consistent with one another.
 
The value of $\tdisref$ can be derived from a comparison between the
predicted mass or age distributions (Eqs. \ref{eq:Nmdisruption} or
\ref{eq:Ntdisruption}) and the observed fits of Table \ref{tbl:M51fits},
or from the values of $\tcross$ or $\Mcross$ (Eq. \ref{eq:Mcross} or
\ref{eq:tcross}). Since \tcross\ and \mcross\ are consistent with each
other, they give the same value of $\tdisref$, but since \tcross\ could be
determined with a higher accuracy than \Mcross\ we use the expression
for \tcross\ to calculate $\tdisref$. Taking into account the uncertainty
in \tcross\ and in $\gamma$ we find from Eq. (\ref{eq:tcross}) that
$\log(\tdisref) = 7.64 \pm 0.22$ in yrs. Combining this with the derived
value of $\gamma$ we conclude that the disruption time of clusters in
the observed area of M51, i.e. the inner spiral arms, is

\begin{equation}
\log \tdis=7.64 (\pm 0.22)+0.57~(\pm 0.10) \times \log (\Mcl /
10^4)
\label{eq:M51tdis}
\end{equation}
with \tdis\ in years and the initial cluster mass in \Msun.
This relation is valid for clusters in the mass range of $10^3$ to 
$10^5$ \Msun. We see that clusters with an initial mass of 
$\Mcl = 10^4$ \Msun\ disrupt in
about 40 Myr and those of $10^5$ \Msun\ in 160 Myr. 
We will show below
that this is much faster than the disruption times of clusters in the
other galaxies that we studied.


\section{The disruption of clusters in M33}
\label{sec:4}

The spiral galaxy M33 is at a distance of 840 kpc (Madore et al. 1991)
and it has an inclination angle of 56 degrees (Regan \& Vogal, 1994).
Chandar et al. (1999a, 1999b, 1999c)  have studied  60 clusters in 
20 $HST-WFPC2$ fields of M33 at galactocentric distances between
0.8 and 5 kpc (with one exception at 0.19 kpc). 
The clusters were observed in the $HST$ $UBV$ and $F170W$ filters.
The limiting magnitude of the sample is $\Vlim \simeq 19.3$.

Chandar et al. (1999b) determined the ages and masses of the clusters
from a study of the energy distributions, by comparing the colours
with those predicted by Bertelli et al. (1994) for instantaneous
star formation with Salpeter's IMF, for three metallicities.
Since there is a gradient in metallicity in M33, the metallicities of
the clusters were derived from optical spectroscopy, using metal
dependent indices defined by Huchra et al. (1996). The 
extinction of the clusters was derived from colour-colour diagrams and
the masses and ages were derived by comparison of the energy
distribution with the predicted ones.  Chandar et al. (1999c)
redetermined the ages and masses from the study of those clusters that
also have a detectable flux in the $F170W$ filter. The combination of
the data of Chandar et al. (1999b, 1999c) provides a list of 49
clusters with an age determination and 45 clusters with a mass
determination. The authors quote an uncertainty in age of about 
0.15 to 0.30 dex for clusters older than $10^8$ years and 0.1 dex for
younger clusters. The derived masses have a quoted accuracy better than
0.1 dex in most cases and better than 0.3 dex in the worst case.
Chandar et al. (1999b, 1999c) argued that the cluster formation rate in 
M33 was about constant.

We use the data from Chandar et al. (1999b, 1999c) to study the
disruption of clusters in M33, in the same way as done for M51.
Metallicity differences will influence the evolution of the stars,
which is taken into account in the fitting of the energy distributions
by Chandar et al. Metallicity is not expected to affect the disruption
of the clusters directly. Therefore we can use the statistics of the clusters,
independent of their metallicity.

\begin{figure}
\centerline{\psfig{figure=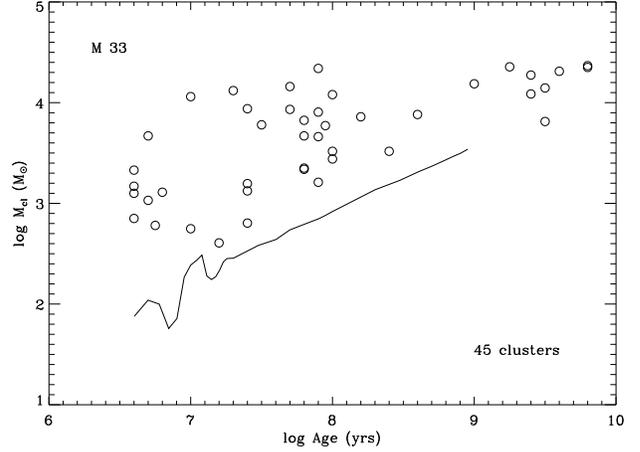, width=9.0cm}}
\caption{The relation between the initial mass \Mcl\ 
(in $M_{\sun}$) and age (in yrs) of 45 M33 clusters studied by
Chandar et al. (1999b, 1999c). The full 
line is the expected lower limit due to fading of aging clusters,
as in Fig. (\protect{\ref{fig:M51mt}}) but shifted vertically
to account for the smaller distance of M33 (viz. 0.84 Mpc instead of
8.4 Mpc) and for the brighter
detection limit of the clusters (viz. $V \simeq 19.3$ instead of $
\simeq$ 22.0)}
\label{fig:M33mt}
 \end{figure}

Figure (\ref{fig:M33mt}) shows the mass versus age relation
of the 45 clusters with
known age and mass. The full line is the magnitude limit for
clusters with solar metallicity. The shape is the same as in Fig. 
(\ref{fig:M51mt}), but the curve is shifted vertically 
by 0.92 dex to allow for the fact that the distance of M33 is ten
times smaller than the distance of M51, and for the fact that the
limiting magnitude of the M33 clusters is 2.7 magn. brighter than for
the M51 clusters. We see the similar
effects as in the distribution of the M51 clusters,
discussed in Sect. \ref{sec:3c}:\\
(a) the observed lower limit shows the expected increase of mass with
age, as predicted for clusters that fade as they get older.\\
(b) the number of clusters drops at ages higher than about
$10^8$ years. The effect is even stronger than it appears in the
figure, because of the logarithmic age scale of the plot.

\

\begin{figure}
\centerline{\psfig{figure=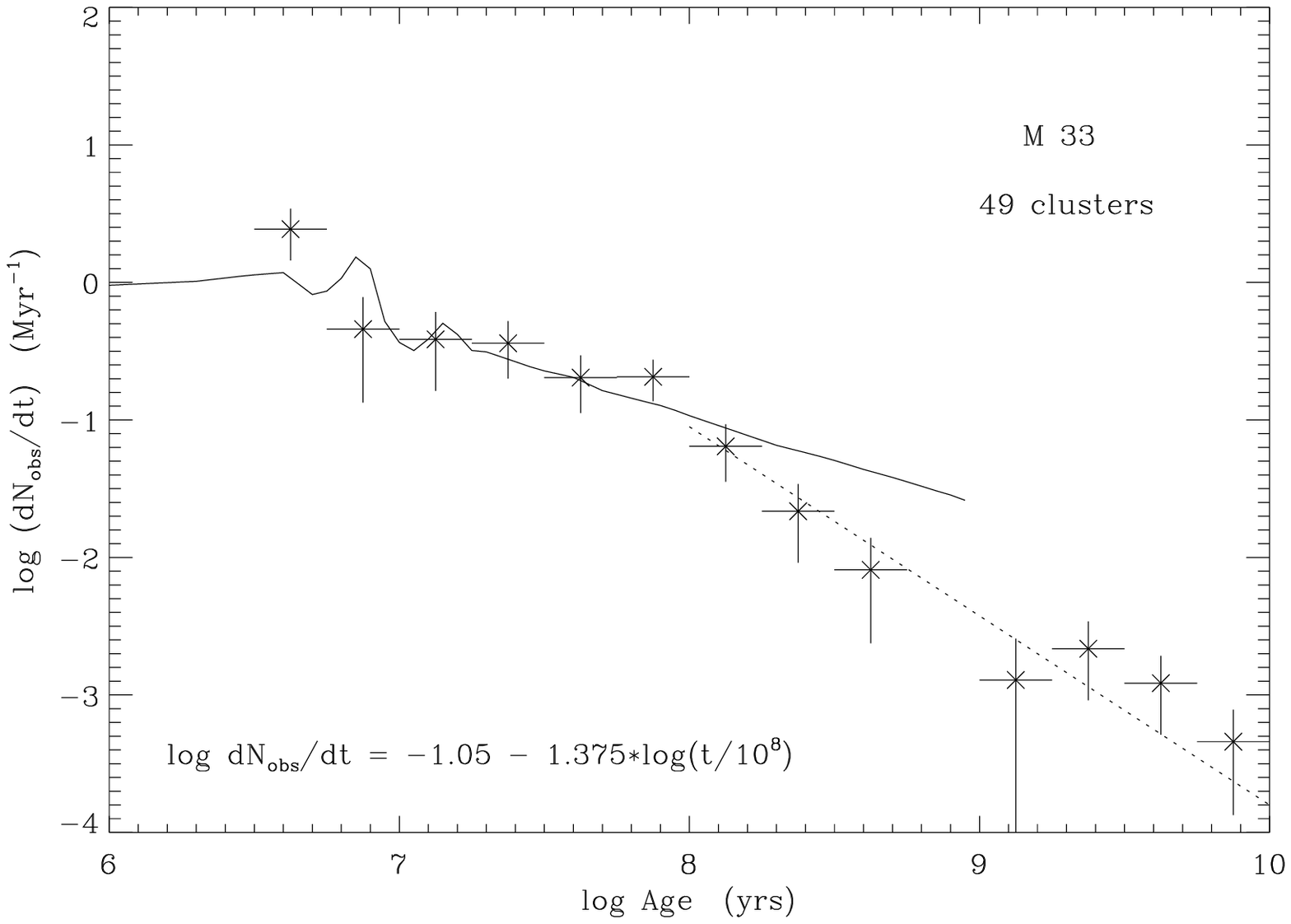,width=9.0cm}}
\centerline{\psfig{figure=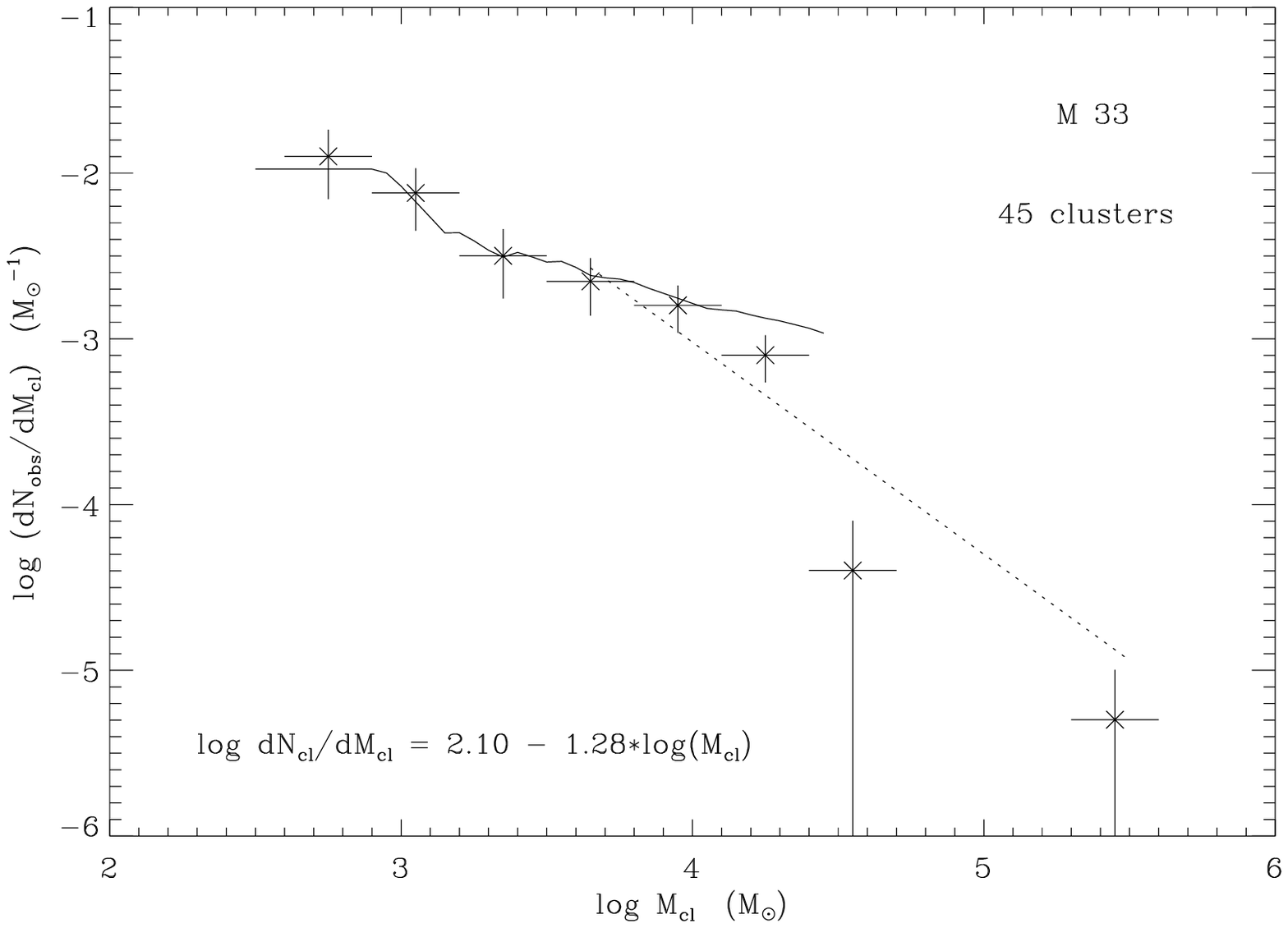,width=9.0cm}}
\caption{The age (upper=Fig a.) and mass (lower=Fig b.) 
distribution of the observed clusters in M33. The distribibutions are
fitted with two lines: the full lines show the expected decrease
due to the detection limit only (fading). 
The dashed line in the upper figure is the least square
powerlaw fit for the clusters of $t>10^8$ years.
The expression of the fit is  given in the figure.
 This line indicates the
dependence of the disruption time on mass.
The dashed line in the lower figure was derived from the 
dashed line in the upper figure (see text).
}
\label{fig:M33dist}
\end{figure}

Figures (\ref{fig:M33dist}a) and (\ref{fig:M33dist}b) 
show the age (top) and mass (bottom)
distribution of 49 and 45 clusters respectively. The full line is the 
prediction due to the fading of the clusters. The mass distribution
follows the prediction for fading clusters up to a mass of about
$2 \times 10^4$ \Msun\ and a steep drop to higher values. The drop is even
steeper than suggested by the figure, because several mass bins contain
no clusters at all. The data for higher masses are too scarce to
derive a disruption relation from the mass distribution.

The age distribution of 49 clusters (Fig. \ref{fig:M33dist}a)
is rather flat for clusters with ages below $10^8$
years. The slope of this distribution agrees roughly with the
predictions for fading clusters, as shown by the full line.
The high point at $\log t = 6.6$ and the low point at 6.9 
(compared to the predicted relation) are 
probaly spurious. They depend on the accuracy of the age determination
of very young clusters, and on the limits of the age-bins used for
this plot. At $\log t \simgreat 8$ the observed distribution 
drops below the predicted relation. This shows that older clusters
must have been disrupted. A least square fit through the data yields 
the empirical relation

\begin{equation}
\log \dd \Ncl /\dd t= -7.05 - 1.38 (\pm 0.18) \times \log (t/10^8)
\label{eq:dndtM33}
\end{equation}
in clusters per yr.
This relation is shown by a dashed line. 

The crossing point is at $\log (\tcross)=7.90 \pm 0.2$.
Adopting a cluster IMF of
$\alpha=2$ we find that the disruption time of the clusters scales
with mass as $\tdis \sim \Mcl^{\gamma}$ with $\gamma = 0.72 \pm 0.12$
(see Eq. (\ref{eq:Ntdisruption})). Using this slope we can predict the
corresponding relation for the mass distribution (see
Eq. (\ref{eq:Nmdisruption})), which results in a predicted slope of
$-1.28 \pm 0.12$.
This relation is shown in Fig. (\ref{fig:M33dist}b) by a dashed line.
The vertical shift of the line is fitted through the last three data points.
We see that this relation fits the data quite poorly. This is due to
the very small numbers of clusters with initial masses above $10^4$
\Msun\ in the sample studied by Chandar et al. (1999b, 1999c).
The crossing point of the (dashed) disruption line and the (full)
fading line in this figure is at $\log (\Mcross) = 3.70\pm 0.3$.
If we had adopted the lower limit of $\gamma=0.60$ the disruption
line would have been slightly steeper, fitting the data slightly
better, but the value of $\Mcross$ would have changed very little.

Converting Eq. (\ref{eq:dndtM33})
into a relation for the cluster disruption time, by using the fitting
of the fading lines to determine the values of 
$\log (S)=-3.24$ and $0.4 (\Cref-V_{\rm lim})=-1.47$ (Eqs. 
(\ref{eq:Nmnodisruption}) and (\ref{eq:Ntnodisruption}))
similar to the method applied in Sect. \ref{sec:3d3}, we find 

\begin{equation}
\log \tdis=8.12 ~(\pm 0.30)+0.72~(\pm 0.12) \times \log (\Mcl / 10^4)
\label{eq:tdisM33}
\end{equation}
for M33 clusters. 
Comparing this expression with the one derived for M51
(Eq. (\ref{eq:M51tdis})) we see that the slope is slightly shallower in
M51, but within the errors of the rather uncertain slope derived for
M33. The constant $\tdisref$ for the disruption time is about a factor 3 times
larger for M33 than for M51.


\section{The disruption of clusters in the SMC}
\label{sec:5}

Hodge (1987) has determined the ages of 326 clusters
in the SMC  from the apparent B-magnitude of
the brightest star in each cluster. 
The accuracy of the age determination is about 40 \%. 
We will use this sample, rather than the more homogeneous sample of 
cluster ages from isochrone fitting by Pietrzynski \& Udalski (1999),
based on the $OGLE-II$ photmotery, because that sample contains only
7 clusters older than 300 Myrs, 4 of which  have an uncertain age. 

The sample of clusters with age determinations by Hodge (1987), 
was determined from photographic plates taken with the
CTIO 4-m telescope, down to a limiting magnitude 
of $B\simeq22$ to 23 (Hodge 1983). This corresponds to an absolute
magnitude  limit of $M_B \simeq +4$ to $+3$, 
or a turn-off age of about 10 to 15 Gyrs. This
implies that clusters of all initial masses, down to a few hundred \Msun\,
should be detectable. In other words, the age distribution of the 
observed clusters is not affected by fading effects described in 
Sect. \ref{sec:2a}, so the value of $\zeta$ in the 
age distribution  is expected to be zero.

There is, however, another selection effect of the sample.
The clusters were identified by Hodge (1983) on the basis of visual
inspection of photographic plates. These clusters have a radius
between 2 and 15 pc, with a mean radius of 7 pc.
Dense clusters are more easily
recognized than less dense clusters, and massive clusters are more
easily recognized than low mass clusters. So, although the magnitude
limit of the brightest star in the cluster does not produce a bias in
the age distribution, the selection of the clusters may be biased
to the denser and more massive clusters. 
This bias is described in Appendix B, where we will show that it is
not important for the SMC clusters studied by Hodge, except
for the oldest ones. For the moment we ignore this 
possible bias in the detection of the SMC clusters.

\begin{figure}
\centerline{\psfig{figure=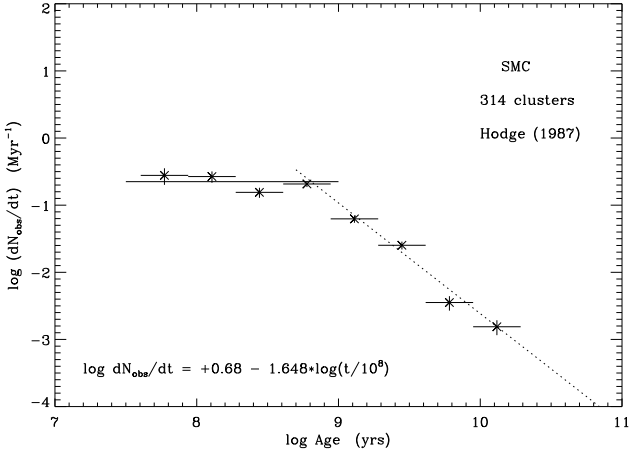,width=9.0cm}}
\caption{The age  
distribution of the clusters in the SMC with age determinations by
Hodge (1987). 
The distribibution is
fitted with two lines: the horizontal full line indicates about a constant
cluster formation rate.
The dashed line is the least square
powerlaw fit for the distribution of older clusters.
The expression of the fit is given in the figure.}
\label{fig:SMC}
\end{figure}

Figure (\ref{fig:SMC}) shows the age distribution of
314 SMC clusters with age determinations by Hodge (1987).
The figure shows a distribution that is flat for ages less than about
1 Gyr and decreases steeply to higher ages. The horizontal
part 
supports the assumption that selection effects and fading play a minor role.
The least square fit to the decreasing
part gives

\begin{equation}
\log \dd \Ncl /\dd t=-5.32 - (1.65 \pm 0.22) \times \log (t / 10^8)
\label{eq:SMCage}
\end{equation}
in clusters per year. The crossing point of the two relations occurs at
$\log t = 8.8 \pm 0.2$, in yrs. 
The slope of the powerlaw fit implies that
$\gamma = 0.61 \pm 0.08$ if $\alpha = 2.0$.

We can estimate $\tdisref$ 
from the crossing point between the horizontal part and the sloped part of
the observed $\log(d \Ncl / dt)$-relation in Fig. (\ref{fig:SMC}).
The horizontal part of the distribution for ages less than
about 1 Gyr, shows that the fading of clusters has played no role
in the selection of the clusters, so the constant $\zeta$
in Eq. (\ref{eq:VMt}) is about zero. In that case the predicted
age distribution for clusters without disruption is

\begin{equation} 
\frac{\dd \Ncl}{\dd t} = \int_{\Mlim (t)}^{\Mcl(\rm max)} S~\Mcl^{-\alpha} ~d \Mcl
\simeq ~\frac{S}{\alpha-1}~ M_{\rm min}^{1-\alpha}
\label{eq:Ntconstant}
\end{equation}
where $ \Mlim = \Mmin$ is the minimum mass of the detected clusters. 
We see that
$d \Ncl / dt$ is predicted to be a constant,
 as is indeed observed for the younger
clusters in the SMC. In case of disruption, the
age distribution is given by Eq. (\ref{eq:Ntdisruption}). The 
crossing point of the two equations is located where

\begin{equation}
\frac{t_{\rm cross}}{\tdisref}~=~(\Mmin/10^4)^{\gamma}
\label{eq:t0}
\end{equation}
The minimum initial mass of the clusters detected by Hodge (1983)
is estimated to be about $\log (\Mmin)\simeq 2.3\pm 0.3$ for the SMC 
(see Appendix B).
The value of $\tdisref$ can now be estimated from
the observed crossing point of the two relations of the 
age distribution of the SMC in Fig. (\ref{fig:SMC}). 
This results in the disruption relation

\begin{equation}
\log \tdis=9.9(\pm 0.2)+0.61~(\pm 0.08) \times \log (\Mcl / 10^4)
\label{eq:tdisSMC}
\end{equation}
We see that the disruption time of clusters in the SMC
is considerably larger than for M51 and M33, but that the slopes of
the disruption relations are very similar.


\section{The disruption of clusters in the solar neighbourhood}
\label{sec:6}

Wielen (1971) studied the age distribution of galactic clusters
from the catalogues of Becker \& Fenkart (1971) and Lindoff (1968).
His samples contain respectively 70 and 59 clusters
within a projected distance of 1 kpc from the sun. The ages of the
clusters studied by Becker \& Fenkart were determined from the
colour-magnitude diagrams, with the age calibration by Barbaro et
al. (1969). The ages of clusters from the Lindoff catalogue were
also derived using the Barbaro et al. isochrones by Wielen.
We will use the clusters from the
Becker \& Fenkart catalogue and from the Lindoff catalogue
with the ages listed by Wielen.  
Wielen (1971) has shown that the above mentioned lists of clusters, within a 
galactic distance of 1 kpc from the Sun, is not complete, but that they
provide  representative 
samples of the age distribution of the clusters. Therefore we will use 
these two samples to derive the disruption of galactic clusters from a
study of their age distribution.

\begin{figure}
\centerline{\psfig{figure=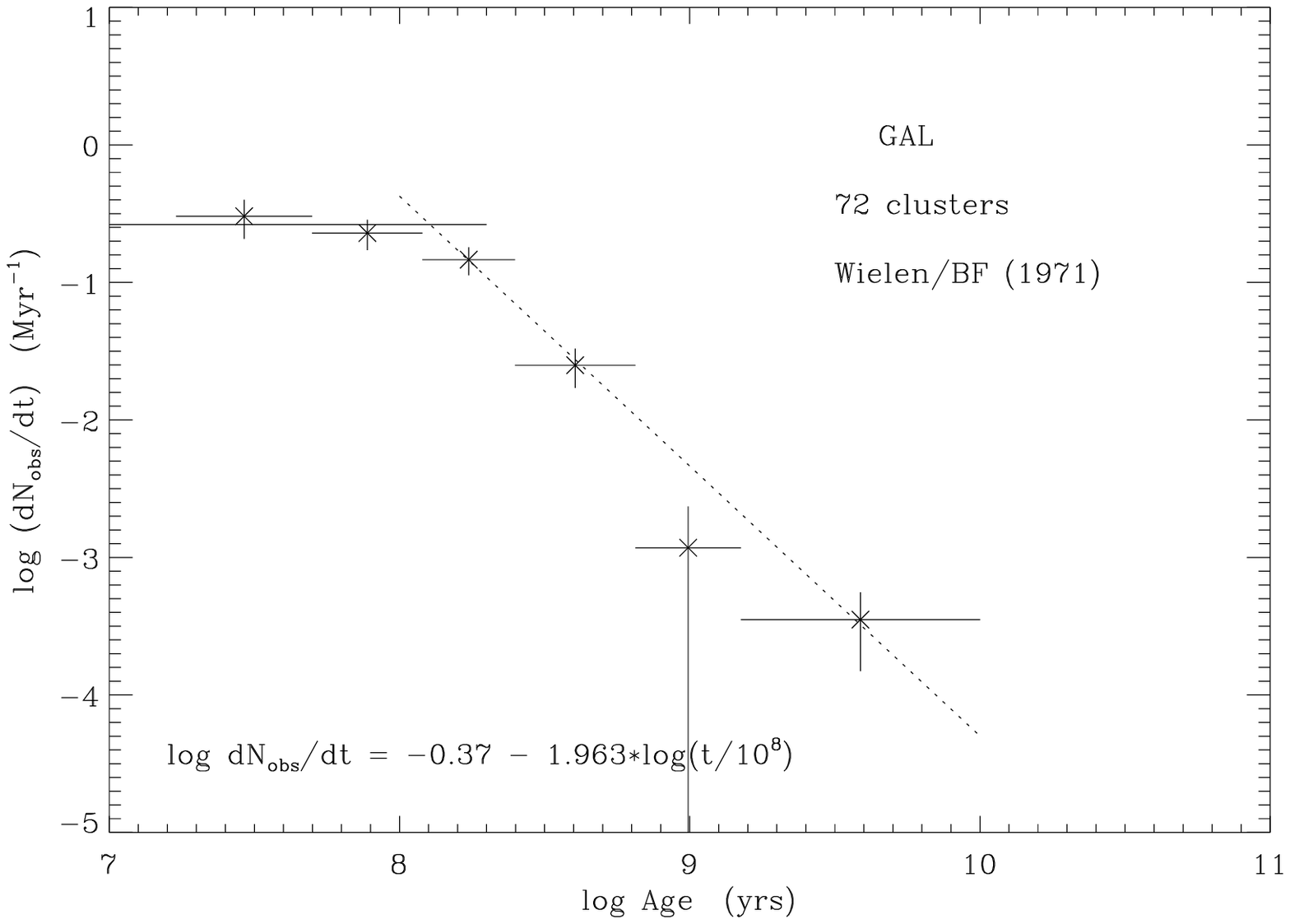,width=9.0cm}}
\centerline{\psfig{figure=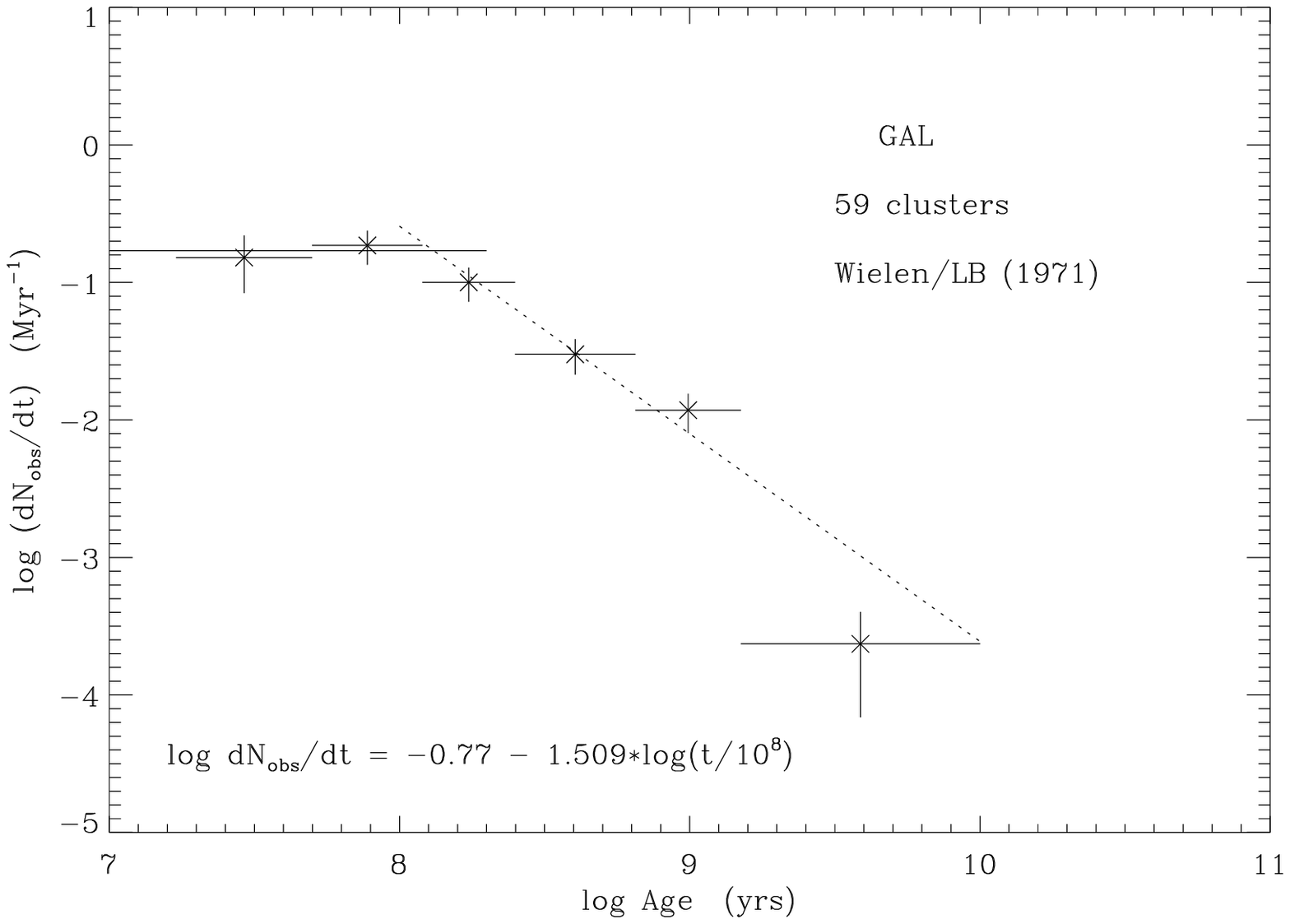,width=9.0cm}}
\caption{The age  
distribution of the observed Galactic clusters 
within a projected distance of 1 kpc from the Sun. 
The upper figure is for the Becker \& Fenkart (1971) sample (BF)
and the lower figure is for the Lindoff (1968) sample (LB).
The distribibutions are
fitted with two lines: the horizontal full lines indicate the constant
cluster formation rate.
The dashed lines are the least square
powerlaw fits for the distributions of older clusters.
The expressions of the fits are given in the figure.}
\label{fig:gal}
\end{figure}

Figures (\ref{fig:gal}a) and (\ref{fig:gal}b) show the age distributions of the
clusters, expressed in $d \Ncl / dt$, of the clusters from the
Becker and Fenkart (1971) catalogue (hereafter called BF)
and from the Lindoff (1968) catalogue with the age calibration by
Barbaro et al. (1969), hereafter called LB. Both distributions
show rather similar characteristics: a flat part at $\log t < 8.0$,
and a steeply declining part for higher ages. The flat part
corresponds to a cluster formation rate of
$10^{-0.58}$ clusters~Myr$^{-1}$ for the BF sample and $10^{-0.77}$ 
clusters~Myr$^{-1}$ for the LB sample. The steep parts are fitted with
linear least square fit relations

\begin{eqnarray}
\log \dd \Ncl /\dd t&=& -6.37 - (1.963 \pm 0.43) \times \log (t / 10^8)
\nonumber \\
\log \dd \Ncl /\dd t&=& -6.77 - (1.592 \pm 0.42) \times \log (t / 10^8) 
\label{eq:Galage}
\end{eqnarray}
in clusters per yr
for the BF and LB samples respectively.
The resulting values of the mass dependence of the disruption law
(Eq. (\ref{eq:tdis})) are respectively 
$ \gamma = 0.59 \pm 0.12$ and $ \gamma=0.63 \pm 0.18$ if the 
CIMF has a slope of $\alpha = 2$.

The almost horizontal distribution for young clusters suggests that there
is no bias towards the more massive ones in the samples of 
Galactic clusters used here, as was already argued by Wielen (1971).
Therefore we can estimate the cluster disruption time from the
crossing points, $\log \tcross = 8.1$, 
of the age distributions shown in  Fig. 
(\ref{fig:gal}), in the same way as applied for SMC sample.

For galactic clusters within 1 kpc we adopt the minimum mass of about 
$ \log \Mmin \simeq 2.5 \pm 0.5~ \Msun$. 
This is about as high as for the SMC, because the
higher brightness of the Galactic cluster stars, compared to the SMC
stars, is partly compensated by the larger angular diameter of the
Galactic clusters, which will make the detection of sparse 
clusters more difficult.
The crossing point then indicates a disruption time of

\begin{equation}
\log \tdis~\simeq ~9.0~(\pm 0.3) +0.60 ~(\pm 0.12) \times \log (\Mcl / 10^4)
\label{eq:tdisgal}
\end{equation}
This disruption time is longer than for the M51 clusters, but shorter
than for the SMC clusters.

\section{Comparing the disruption times of different galaxies}
\label{sec:7}

We have derived estimates of the parameters $\tdisref$ and $\gamma$ that describe 
the dependence of the cluster disruption
time on the {\it initial} cluster mass, Eq. (\ref{eq:tdis}), for
different galaxies. These parameters are summarized in
Table (\ref{tbl:1}). Columns 2 and 3 give the number of clusters used for
the determination of the disruption times. Columns 4 and 5 give the
age range and the mass range of the clusters.
Column 6 gives the disruption
time $\tdisref$ of a cluster with initial mass of $10^4$ \Msun. Column 7 gives
the derived slopes of the age distribution (expressed in
$\log \dd \Ncl / \dd t$). Column 8 gives the resulting slope of the age
dependence of the disruption law, {\it under the assumption that the
  cluster IMF has a slope of $\alpha=2$.}

\begin{table*}
\caption[]{ The parameters of the disruption time:
~~$\tdis = \tdisref \times (\Mcl/ 10^4)^{\gamma}$}
\begin{tabular}{lccccccc}
\hline \\
Galaxy & Nr  & Nr &  Age range & Mass range & $\log \tdisref$  & 
$(1-\alpha)/\gamma$ & $\gamma$ \\
       &   $\log d\Ncl /dt$& $\log d\Ncl/ d M$ &log (yrs) & 
log (\Msun) & (yrs)  &     &    \\
\hline \\
 M51       &  380/512/744 & 380/512/744 & 6.0 -- ~9.7  & 3.0 -- 5.2 & 
$7.64 \pm 0.22$ & $1.75 \pm 0.32$ & $0.57 \pm 0.10$ \\
 M33       &  49 & 45     & 6.5 -- 10.0 & 3.6 -- 5.6 &
$8.12 \pm 0.30$ & $1.38 \pm 0.18$ & $0.72 \pm 0.12$ \\
 Milky Way & 72/59& ---   & 7.2 -- 10.0 &     ---    &
$9.0 \pm 0.3$   & $1.77 \pm 0.40$ & $0.60 \pm 0.12$ \\
 SMC       &  314& ---    & 7.6 -- 10.0 &     ---    &
$9.9 \pm 0.2$   & $1.65 \pm 0.22$ & $0.61 \pm 0.08$ \\
 & & & & & & Mean & $0.62 \pm 0.06$ \\
\hline\\
\end{tabular}
\label{tbl:1}
\end{table*}

The values of $\gamma$ are very similar (within the uncertainties)
for the different galaxies. The mean and its standard deviation is

\begin{equation}
\gamma~=~ 0.62 \pm 0.06
\label{eq:gamma}
\end{equation}
The values of $\tdisref$, however, are very different for the different
galaxies: M51 has the shortest cluster disruption time and the
SMC has the longest disruption time. The difference amounts to
about a factor $10^2$.


\section{Discussion}
\label{sec:8}


\subsection{The description of the disruption time}
\label{sec:8a}

In our simple description of the disruption time 
as a function of the initial mass of the cluster (Eq. (\ref{eq:tdis}))
we have assumed that 
for each initial mass the cluster stays intact up to a certain
time \tdis\ and then suddenly disappears. This is of course a
severe simplification of the true situation.

  Theoretical and dynamical simulations (e.g. Spitzer 1957; 
  Portegies Zwart et al. 1999; Fall \& Zhang 2001)
  both suggest that
  disruption results in a linear decrease of the mass of a cluster
  with time, rather than the sudden disappearance that we adopted 
  in this study. Therefore our assumption of sudden disruption is a
  severe simplification.

 However, it can be shown that, even for
  gradually disrupting  clusters, the logarithmic age and mass
  histograms of the surviving clusters above the detection limit
  will show a distribution that can be fitted with two powerlaws.
  One of these powerlaws will depend on the fading of the clusters and
  the other one will depend on the cluster decay time, very similar to
  the histograms predicted for sudden disruption, shown in Fig.
  (\ref{fig:Npred}). This is shown in Appendix A, where we also show 
  that the predicted mass versus age
  distributions for gradual disruption are very similar to the 
  observed distributions of clusters in M51 and M33,
  Figs. (\ref{fig:M51mt}) and (\ref{fig:M33mt}).

Theoretical arguments and studies of cluster disruption,
e.g Spitzer (1957), Chernoff \& Weinberg (1990), 
de la Fuente Marcos (1997), Portegies-Zwart et al. (1998)
and Fall \& Zhang (2001),
 show that the disruption time is expected to
depend on their initial  density and internal 
velocity disruption, rather than on their mass,
as was adopted in this study. However, since the radius of most 
of the clusters is not known, we have used their initial mass as 
the only parameter for the description of the disruption time. 
{\bf Basically this implies that we have included the density dependence
of the disruption time in the mass dependence. If the density of
a cluster in a particular sample depends on its initial mass,
e.g. $\rho \sim  M^x$ then the derived value of $\gamma$ reflects 
the dependence of \tdis\ on a
combination of the mass and density dependence, as described in Sect 2.1.
In particular, if the disruption time is proportional to the
relaxation time $\tdis \sim  t_{\rm rxt} \sim M \rho^{-1/2}$, then the 
derived value of $\gamma \simeq 0.62 = 1-x/2$. So our result
is consistent with a mean density-versus-mass relation of $\rho \sim M^{0.76}$.
The empirical determination of the dependence of the disruption
time on both the mass and radius for clusters in M81 
is described in a forthcoming paper.}

\subsection{Continuous cluster formation?}
\label{sec:8b}

In the determination of the disruption rates 
based on the mass and age distributions of the clusters,
we have assumed that the cluster formation rate was constant.
This is probably a reasonable assumption for the solar neighbourhood,
and for M33. However this is not a trivial assumption for 
the interacting galaxy M51 with its companion NGC 5195, nor for the  
SMC which interacts with the LMC and the Galaxy.  

Scuderi et al. (2002) and Lamers et al. (2002) found that the 
nucleus of M51 has a strong star burst with an age of several $10^8$
years, in agreement with the time of closest approach of the
companion. 
Bik02 has studied the age distribution of the M51
clusters in the inner spiral arms  and found only a faint
indication (less than 2 $\sigma$) of an increased cluster formation 
rate in M51 a few $10^8$ years in this region.
This indication can also be seen in the possibly small bump
in the age distribution of Fig. \ref{fig:M51dist}b near
$\log t\simeq 8.7$. If indeed present, this bump  might be resonsible
for the difference between the two values of $\gamma$ derived from the
mass and the age distributions. Ignoring this bump would have
resulted in a slightly smaller vale of $\gamma$ derived from the
age distribution. The effect on the determination of $\tdisref$
from the mass and age distributions together is negligible.

The problem might also exist for the SMC clusters.
Several authors, e.g. Mateo (1988), Bica et al. (1996), Pietrzynski \& Udalski (1999),
(see the review by Da Costa (2002) and references therein), have shown 
that the age distribution of clusters in the LMC shows peaks due to
the interaction with the SMC and the Galaxy. However, the age
distribution of the SMC clusters is much more homogeneous (da Costa
2002). Therefore we did not study the cluster disruption in the LMC
and we 
adopted a constant cluster formation rate 
for the SMC as a first approximation to derive the disruption time 
of SMC clusters. When larger samples of clusters with more accurate
{\it ages and masses} become available, it is possible to verify the
influence of non-constant cluster formation rates on the determination
of the disruption times by consisdering both the mass and the age
distributions.
In case of a non-constant
cluster formation rate, the determination of the 
disruption time can then easily be corrected by a numerical
calculation of the integrals of Eqs. (\ref{eq:Ntnodisruption})
and (\ref{eq:Ntdisruption}) for an age dependent $S$.


\subsection{Comparison with other studies}
\label{sec:8c}

Our resulting values for the parameters of the cluster disruption
relations, summarized in Table (\ref{tbl:1}), can be compared with 
the results of other studies, e.g. by  Elson \& Fall
(1985, 1988), Hodge (1987) and Mateo (1988). 
In these studies the age distributions for clusters in
the LMC, SMC and solar neighbourhood, normalized to the same value at
$t \simeq 100$ Myrs, are compared. The comparisons show clearly that 
the age distribution is steeper for the solar neighbourhood than for
the LMC and SMC. The difference was expressed in terms of a ``mean
age'', $\tau_m$, of the sample of clusters above the detection limit.
Elson and Fall (1985, 1988) found $\tau_m = 2 \times 10^8$ yrs for the solar
neighbourhood and $\tau_m = 4 \times 10^9$ yrs for the LMC, which
indicates that the disruption of clusters in the LMC is
slower than in the solar neighbourhood. 

To compare the results of Elson and Fall (1985, 1988) with our
results, let us assume for the moment that the disruption time
of clusters in the LMC is about the same as the disruption time
of clusters in the SMC, which we have measured.
The factor 20 difference in the mean ages of the LMC and Galactic
clusters,
found by Elson \& Fall appears to be much larger than the 
factor 4 or so in our values of $\tdisref$ between the SMC and the Milky Way. 
However, one should bear in mind that even a small difference in $\tdisref$
can produce a large difference in the mean age of the observed
clusters. This is because the mean age depends on the mass 
and age range of the studied clusters, and their brightness above the
detection limit. In fact, we used the same sample of SMC and
Galactic
clusters as Hodge (1988) and almost the same sample as used by 
Elson \& Fall (1985). So the difference of only a factor 4
in $\tdisref$ between the SMC and the Galaxy is consistent with 
the factor 20 difference in mean cluster age. Our analysis has the
advantage that it results in a description of the mass dependence of
the disruption times in both galaxies. 

For the solar neighbourhood we compare our result with that of
van den Bergh (1981).
The disruption time of 1 Gyr for a $10^4$ \Msun\ cluster can be
compared with the e-folding time 0.15 Gyr for the disruption of
clusters within 0.75 kpc derived by van den Bergh (1981), who
used the sample of 63 clusters from the catalogue by Mermilliod 
(1980). These clusters have a mean $M_V$ of about -3 to -4, which
corresponds to a total mass of  $\log (M/\Msun) \simeq 2.5 \pm 0.2$ 
if the clusters have a mean age of 100 Myrs and 0.3 dex lower if
their mean age is 30 Myrs (Leitherer et al. 1999). From
Eq. (\ref{eq:tdisgal}) we see that our analysis suggests a disruption
time of 0.13 Gyr for clusters of $M \simeq 300~
\Msun$. This agrees
surprisingly well with the mean value of 0.15 Gyr 
derived derived by van den Bergh
(1981), considering the uncertainties involved.

 
 \subsection{Possible biases}
\label{sec:8d}

The low extinction
values of the clusters in M51 studied by Bik et al. (2002) and 
in M33 studied by Chandar et
al. (1999b, 1999c), due to the 
almost face-on orientations of M51 and M33, together with the fact
that the clusters were detected with one instrument under 
constant conditions, results in a minimum bias in the statistics 
of these cluster samples.
The fading of the clusters due to the aging of the stars,
makes the older low mass clusters drop below the detection limit.
However the resulting bias can very well be taken into account
because the cluster models predict the fading of the clusters in
the different wavelength bands as a function of age and cluster mass.
The predicted bias in the age and mass statistics of the clusters
is nicely confirmed by the statistics of the observed clusters with ages
younger than the disruption times (see Figs.
(\ref{fig:M51mt}) and (\ref{fig:M33mt})).

For the studies of clusters in the SMC 
this method of correcting for the expected fading does not apply, 
because the clusters were selected by visual inspection of
photographic plates (Hodge 1986, 1988). For the 
SMC clusters we have assumed that the sample is incomplete but 
relatively unbiased down to some minimum cluster mass, because the stars
above a certain brightness limit could be detected individually
(see Sect. \ref{sec:5} and Appendix B). The nearly flat part in the
age distribution of the SMC cluster sample
(Fig. (\ref{fig:SMC})) confirms that this was a reasonable
assumption.

For the clusters in the solar neighbourhood the main
bias is due to the variable large extinction. However, Wielen (1971)
has described several tests that show that, although the sample within a
distance of 1 kpc may be incomplete, it is relatively free of bias.
Also in this case we have estimated the lower mass limit of the
detected clusters in an approximate way, similar to the method applied
to the SMC clusters. Again the flat part of the age
distribution for ages less than the disruption time supports this
assumption. Therefore we feel confident that the disruption time can
indeed be derived from the cluster samples of the SMC and solar
neighboorhood. 

The situation can be improved 
when the detection of clusters can be made in an
automatic and unbiased way (or rather, with a bias that is known and
can be corrected for) from large homogeneous data bases.


\section{Summary}
\label{sec:9}

 Under the assuptions that:\\
(a) clusters are formed at a constant rate with the same cluster IMF
and the same stellar IMF, and \\
(b) that clusters disrupt suddenly after a
  certain time $\tdis(M_i)$ that depends as a powerlaw on their initial
  mass $M_i$,\\
 we derived  the following results.

\begin{enumerate}

\item{} Both the mass distribution and the age distribution of 
clusters in M51 and M33 show a
double powerlaw distribution, with a small slope for young and low
mass clusters
and a steeper slope for older and more massive clusters.
\item{} Such a distribution is predicted for clusters that
are formed with a single CIMF, but
fade due to stellar evolution below the detection limit 
and disrupt suddenly at a time that depends 
as a powerlaw on their initial mass, 
$\tdis = \tdisref \times (\Mcl/ 10^4 \Msun)^{\gamma}$. 

\item{} The age distribution of clusters in the SMC and the solar
neighbourhood also shows this characteristic double powerlaw
distribution, with a flat age distribution for small ages.
This is in agreement with the fact that evolutionary fading 
below the detection limit is not
important for the young observed clusters in these galaxies.
For these galaxies we do not know the mass distribution of the 
used cluster samples.

\item{} From the slopes and the crossing points of the powerlaws 
of the age and mass distributions
the values of $\tdisref$ and $\gamma$ are derived. For M51 the
results can be checked because the slopes and crossing points of the
age distribution and the mass distribution should be consistent with
one another. For the M33, SMC and the solar neighbourhood, the lack of a
reliable mass distribution prevents this check.

\item{} The value of $\gamma$ is
the same for the four galaxies within the uncertainty of the
observations, with a mean value of 
 $\gamma = 0.62 \pm 0.06$ if the CIMF has a slope of
$\alpha=2$, i.e. $\dd N_{\rm cl}/\dd \Mcl \sim \Mcl^{-\alpha}$. If $\alpha > 2$
then $\gamma <0.60$. For other values of $\alpha$ the value of
$\gamma$ has to be corrected by 
$\Delta \gamma \simeq -0.6 \Delta \alpha$.

{\bf
\item{} If the disruption time of the cluster samples that we have
  studied here is proportional to their relaxation timescale,
  $\tdis \sim t_{\rm rxt} \sim M \rho^{-1/2}$, then the derived 
value of $\gamma \simeq 0.62 \pm 0.06$ suggests a mass-density relation
of the type $\rho \sim M^{x}$ with $x \simeq 0.75 \pm 0.10$.
}

\item{} The constant $\tdisref$, which is the disruption time of a cluster of
$10^4$ \Msun, differs drastically from galaxy to galaxy, between 
40 Myrs for clusters in M51 at galactocentric distances of
$0.8 < r< 3.1$ kpc, and 8 Gyrs for the SMC. The disruption time  $\tdisref$ is
about 130 Myrs for the region at galactocentric distances of about
0.8 to 5 kpc in M33 and 1.0 Gyr for the solar neighbourhood.

\item{} The masses of the clusters were derived from a comparison
of their observed energy distributions with Starburst99 models
(Leitherer et al. 1999). These cluster models have a stellar
lower mass limit of 1 $\Msun$. If the true lower mass limit $M_{\rm low}$
is smaller, 
the cluster masses are underestimated by a factor 2.09 if $M_{\rm
  low}=0.2~ \Msun$. 
In that case the value of $\tdisref$ has  to be decreased by about
0.18 dex.

\end{enumerate}



\section*{Acknowledgements}

We are very grateful to Arjan Bik and Nate Bastian for their 
determination of the photometry of clusters in M51. We thank Rupali Chandar 
for providing the cluster data of M33.
We  benefitted from stimulating discussions with Mike
Fall and Simon Portegies Zwart.
We thank Nate Bastian, Mike Fall, Jun Makino and Brad Whitmore 
for constructive comments and suggestions on drafts of this paper.
We are grateful to an unknown referee for questions, suggestions and comments
that improved the quality of this paper.
Stratos Boutloukos is thankful to the Astronomical Institute at Utrecht for
hospitality during this study. His stay was supported by a grant from
the Erasmus project of the European Community.



\section{Appendix A: comparison between instantaneous and gradual 
disruption }

In the analysis of the mass and age distributions  
we assumed that clusters disrupt instantaneously. Theory predicts that
clusters are in fact disrupted gradually. In this appendix 
we will show that gradual disruption produces almost
the same mass and age distributions as instantaneous disruption. 

Suppose that clusters form at a rate and with a cluster initial mass
function given by Eq. (\ref{eq:CIMF}).
Suppose that the disruption time scales with mass as 
given by Eq. (\ref{eq:tdis}).
Then the mass of a cluster decreases with time as 
$\dd M/\dd t \simeq -M/\tdis$ and so

\begin{equation}
M^{\gamma}=M_i^{\gamma}-\gamma 10^{4 \gamma}\frac{t}{\tdisref}=
M_i^{\gamma}-B~t
\label{eq:AppAMiM}
\end{equation}
where $\Mi$ is the initial cluster mass (in \Msun) 
and $B=\gamma 10^{4 \gamma}/\tdisref$.
Suppose that the detection limit of clusters with a present mass
$M$ and an age $t$ is given by

\begin{equation}
M_{\rm lim}=M_L (t/t_L)^{\zeta}
\label{eq:AppA2}
\end{equation}
This expression is equivalent to Eq. (\ref{eq:Mlim}).
Then it is easy to show that  clusters with a {\it present mass} $M$
can be observed if their age is in the range of

\begin{eqnarray}
t_{\rm lower}& =& \frac{\Mmin^{\gamma}-M^{\gamma}}{B}
< t < \nonumber \\
& & {\rm min} \left[ \frac{\Mmax^{\gamma}-M^{\gamma}}{B}; 
t_L\left(\frac{M}{M_L}\right)^{1/\zeta} \right]= t_{\rm upper}.
\label{eq:AppAtrange}
\end{eqnarray}
Similarly, clusters of age $t$ can be observed if their {\it present mass}
is in the range of

\begin{eqnarray}
M_{\rm lower}&=&
{\rm max} \left[ M_L\left(\frac{t}{t_L}\right)^{\zeta};
\left(M_{\rm min}^{\gamma}-Bt\right)^{1/\gamma} \right]
< M < \nonumber \\
& & 
\left(M_{\rm max}^{\gamma}-Bt\right)^{1/\gamma} =M_{\rm upper}
\label{eq:AppAMrange}
\end{eqnarray}
The age distribution of the clusters above the detection limit is

\begin{eqnarray} 
& & \frac{\dd \Ncl}{\dd t} = \int_{M_{\rm lower}(t)}^{M_{\rm upper}}
S~M_i(M,t)^{-\alpha} ~d M~ =
\nonumber \\
& & 
\frac{S}{\alpha-1}
\left\{ 
\left(M_{\rm lower}^{\gamma}+B~t\right)^{(1-\alpha)/\gamma} -
\left(M_{\rm upper}^{\gamma}+B~t\right)^{(1-\alpha)/\gamma} 
\right\}
\label{eq:AppANt}
\end{eqnarray}
and the mass distribution is

\begin{eqnarray} 
& & \frac{\dd \Ncl}{\dd \Mcl} = \int_{t_{\rm lower}}^{t_{\rm upper}}
S~M_i(M,t)^{-\alpha} ~d t \nonumber \\
&=& 
\frac{S~t_0}{(\alpha-1)M_0} \left(\frac{M}{M_0}\right)^{\gamma-1} 
\nonumber \\
&\times& \left[
 \left(M^{\gamma}+B~t_{\rm lower}\right)^{(1-\alpha)/\gamma}-
 \left(M^{\gamma}+B~t_{\rm upper}\right)^{(1-\alpha)/\gamma}
\right]
\label{eq:AppANm}
\end{eqnarray}
with the values of $M_{\rm lower}$, $M_{\rm upper}$,
$t_{\rm lower}$ and  $t_{\rm upper}$ given above.

\begin{figure}
\centerline{\psfig{figure=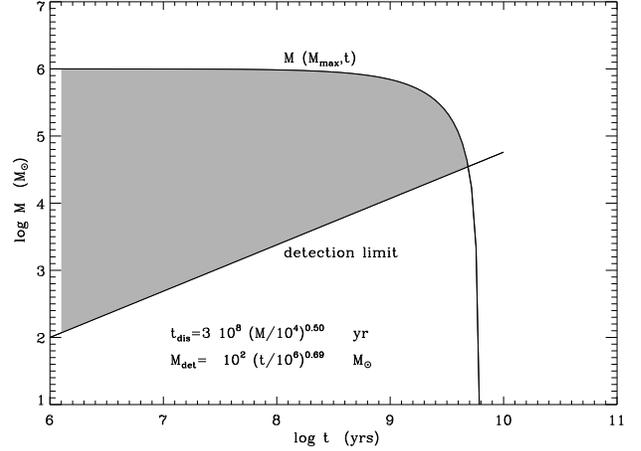, width=9.0cm}}
\centerline{\psfig{figure=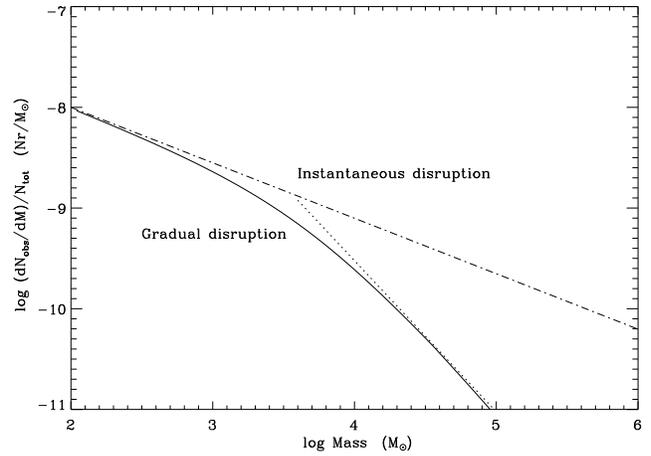, width=9.0cm}}
\centerline{\psfig{figure=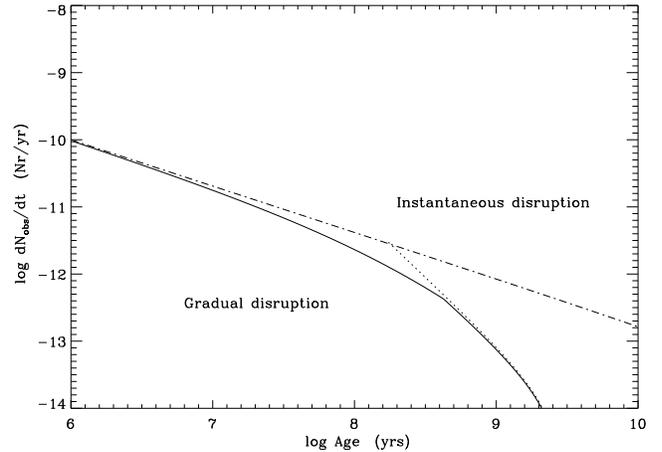, width=9.0cm}}
\caption{Predictions for slow decay of clusters with an age dependent
detection limit.
Upper figure: the predicted distribution of clusters in the mass age
diagram. Middle figure: the predicted mass distribution (full line).
Lower figure: the predicted age distribution (full line).
The dash-dotted lines show the prediction if there was no disruption
but only evolutionary fading. The dotted lines show the prediction
for evolutionary fading and instantaneous disruption.}
\label{fig:appa}
\end{figure}

We have calculated the location of the observable clusters in the
$(M,t)$-diagram and the predicted mass and age distributions of the
observable clusters for a cluster IMF of $\alpha=2$ with 
$M_{\rm min}=10^2$ \Msun\ and $M_{\rm max}=10^6$ \Msun,
for a disruption time with $\gamma=0.50$ and  
$\tdisref=3 \times 10^8$ yrs. 
Figure (\ref{fig:appa})a shows the location of the clusters in the
$(M,t)$-diagram. We see that the clusters occupy an approximately
triangular region 
with the upper limit given by the decrease of $M_{\rm max}$ with age.
The lower limit consists of the detection limit, which is the
lower mass limit that increases with age due to the
evolutionary fading of the cluster. The observed mass versus age 
distributions of the clusters observed in M51 (Fig. \ref{fig:M51mt})
and M33 (Fig. \ref{fig:M33mt}) show the same overall structure.

Figures (\ref{fig:appa})b and c show the age and mass 
distributions of the clusters, normalized to the total number of
clusters formed over a period of 10 Gyrs.  
The full line is the distribution in case of gradual
disruption. The dash-dotted lines give the expected distributions
if only evolutionary fading of the
clusters below the detection limit was important, i.e. no disruption. 
The dotted lines give the distributions in case of  
instantaneous disruption.  In both diagrams we see
that the distributions for gradual disruption approaches the
linear relation of the evolutionary fading at the low mass or age end
and the steeper linear relation for instantaneous disruption at the
high mass or age end, with a smooth transition in between these linear
parts. This demonstrates that, if the observed mass and age distributions are
fitted by a combination of two linear relations (as was done in this
paper) the correct values of the disruption time $\tdisref$ and the 
exponent $\gamma$ can be derived.

\section{Appendix B: the minimum initial mass of detected SMC
  clusters}

The estimate of the cluster disruption time for the SMC
depends on the minimum {\it initial} mass of the clusters detected by
Hodge (1983, 1986, 1987). In this appendix we derive a crude estimate
of this initial mass.

The clusters were detected on blue photographic plates 
of the 4m CTIO telescope (Hodge, 1986).
Stars down to $B \simeq 22$ to $23$ 
could be detected in the SMC. Adopting a distance modulus of 18.85
(van den Bergh 1999) and a mean $E(B-V)=0.08$ or $A_B=0.33$ 
(Pietrzynski \& Udalski 2000) we find $M_B \simeq +3.7 \pm 0.5$
corresponding to a main sequence type between F6 with a mass of about  
1.3 \Msun\ and F1 with a mass of 1.7 \Msun\ (Lang, 1992). So the
minimum mass limit is $1.5 \pm 0.2$ \Msun.

Hodge (Private Communication) estimates that he could 
detect a cluster if it contained
at least between about 10 and 30 stars above the detecting limit,
although some more compact clusters were detected with even fewer stars.
Suppose that the clusters could be detected if they
contained at least 20 stars above these detection limits,
and that the stellar IMF of the clusters can be written as
$N(M)\dd M=C~M^{-2.35}$, i.e. with Salpeters' exponent. The initial mass
of the cluster is 

\begin{equation}
\Mcl = \int_{\Mmin}^{\Mmax} C_{\rm rich} ~M^{-1.35} \dd M 
\simeq \frac{C_{\rm rich}}{0.35} ~ \Mmin^{-0.35}
\label{eq:initialmass}
\end{equation}
where $C_{\rm rich}$ is a parameter that describes the richness of the
cluster. For the minimum mass of the cluster stars we adopt a value of
0.5 \Msun. We assumed that the maximum mass of the cluster stars,
$\Mmax \simgreat 50 \Msun$, is much higher than \Mmin.
The requirement that the detectable clusters should contain at least 
$N=20 \pm 10$ stars brighter than a main sequence star of $1.5 \pm 0.2$
\Msun\ implies that

\begin{equation}
N = \int_{\Mlim}^{\Mmax} C_{\rm rich} ~M^{-2.35} \dd M 
\simeq \frac{C_{\rm rich}}{1.35} ~ \Mlim^{-1.35} > 20
\label{eq:nrstars}
\end{equation}
where $\Mlim=1.5 \pm 0.2~ \Msun$. We find that $C_{\rm rich} \simeq 50
\pm 30$, and
the minimum initial mass of the detectable clusters is $\Mcl \simeq
170 \pm 100~\Msun$. For an adopted minimum stellar mass of $0.25~\Msun$ 
the initial mass is 27 percent larger or $220 \pm 120$ \Msun\
for the detectable clusters.

 In this estimate we have assumed that
the fraction of the stars that have disappeared as supernovae or
dropped below the detection limit as white dwarfs is small compared to
the observable number of stars with masses above 1.4 \Msun.
This is a reasonable assumption for clusters with the
turn-off point above a few \Msun\ and turn-off ages shorter than about 
$10^9$ years. These were the clusters that were used to
derive the value of \tdisref\ in Sect \ref{sec:6}. 
For a conservative estimate we adopt a minimum mass of the detectable
clusters of about $\log(\Mcl)=2.3 \pm 0.3$ \Msun.
 
\end{document}